%% file: ms.tex
\documentclass[letterpaper,twocolumn,10pt]{article}
\usepackage{usenix,epsfig,endnotes}

\usepackage{color}
\usepackage{comment}
\usepackage{multirow,bigdelim}
\usepackage{footnote}
\usepackage{times}
\usepackage{fullpage}

\usepackage[small, compact]{titlesec}
\usepackage{titling}

\usepackage{verbatim} % For comment
\usepackage{xspace} % For xspace
\usepackage[hyphens]{url}
\usepackage[pdftex,
            colorlinks,
            pdfstartview=FitH,
            linkcolor=blue,
            citecolor=black,
            urlcolor=blue,
            filecolor=black
            ]{hyperref} % For refs and urls

\usepackage[numbers,sort&compress]{natbib}
\usepackage{paralist}
\usepackage{enumitem}
\usepackage{subcaption}
\usepackage{booktabs}
\usepackage{tabularx}
\usepackage{fancyvrb}
\usepackage{multirow}
\usepackage{float}
\usepackage[usestackEOL]{stackengine}
\newcolumntype{P}[1]{>{\centering\arraybackslash}p{#1}}

\RequirePackage{xcolor}
\definecolor{RED}{rgb}{1,0,0}
\definecolor{BLUE}{rgb}{0,0,1}
\definecolor{White}{rgb}{1,1,1}

\usepackage{ifthen}
\newboolean{publicversion}
\setboolean{publicversion}{false}
\ifthenelse{\boolean{publicversion}}{
  \newcommand{\grumbler}[3]{}
}{
  \newcommand{\grumbler}[3]{\xspace\textcolor{#3}{\bf #1: #2}}
}

\newcommand{\ra}[1]{\renewcommand{\arraystretch}{#1}}
\newcolumntype{R}[1]{>{\raggedleft\arraybackslash}p{#1}}

\input{macros}

\makeatletter
\newcommand{\printfnsymbol}[1]{%
  \textsuperscript{\hspace{-3pt}\@fnsymbol{#1}}%
}
\makeatother

\begin{document}

\title{ \Large \bf Finding Crash-Consistency Bugs with Bounded
  Black-Box Crash Testing}
%\author{Paper \# 163}

\author{
{\rm Jayashree Mohan\thanks{Both authors contributed equally} \hspace{1pt} \affmark[1]}
\enspace 
{\rm Ashlie Martinez \printfnsymbol{1} \affmark[1]}
\enspace 
{\rm Soujanya Ponnapalli\affmark[1]}
\enspace
{\rm Pandian Raju\affmark[1]}\\
\enspace 
{\rm Vijay Chidambaram\affmark[1]$^,$\affmark[2]}
\vspace{.5em}  \\
\affaddr{\affmark[1]University of Texas at Austin}\hspace{10pt}
\affaddr{\affmark[2]VMware Research}\\
\vspace{-0.4in}
}% end author

\date{}
\maketitle
\input{abstract}
\input{intro}
\input{mot2}

\input{study}
\input{ap4}
\input{tools3}

\input{eval2}

\input{related}
\input{conc}
\input{ack}
\input{appendix}

\bibliography{all,all-confs}
\bibliographystyle{abbrv} 

\end{document}

%% file: macros.tex
\newcommand{\mycaption}[2]{\caption{\textbf{#1}. {#2}}}
\newcommand{\sref}[1]{\S\ref{#1}}
\newcommand{\vheading}[1]{\vspace{0.05in}\noindent\textbf{#1}}

\newcommand{\fsync}{\texttt{fsync()}\xspace}
\newcommand{\etal}{\textit{et al.}\xspace}

\newcommand{\eg}{\textit{e.g.,}\xspace}
\newcommand{\etc}{\textit{etc.}\xspace}
\newcommand*{\affaddr}[1]{#1} % No op here. Customize it for different styles.
\newcommand*{\affmark}[1][*]{\textsuperscript{#1}}

\newcommand{\sysdd}{\texttt{dd}\xspace}

\newcommand{\msync}{\texttt{msync()}\xspace}
\newcommand{\sysopen}{\texttt{open()}\xspace}
\newcommand{\sysclose}{\texttt{close()}\xspace}
\newcommand{\sysrename}{\texttt{rename()}\xspace}
\newcommand{\sysfsync}{\texttt{fsync()}\xspace}
\newcommand{\sysfdsync}{\texttt{fdatasync()}\xspace}
\newcommand{\syswrite}{\texttt{write()}\xspace}

\newcommand{\sysfalloc}{\texttt{fallocate()}\xspace}

\newcommand{\kpsize}{\texttt{KEEP\_SIZE}\xspace}
\newcommand{\fzero}{\texttt{ZERO\_RANGE}\xspace}
\newcommand{\fpunch}{\texttt{PUNCH\_HOLE}\xspace}

\newcommand{\syslink}{\texttt{link()}\xspace}

\newcommand{\sysunlink}{\texttt{unlink()}\xspace}

\newcommand{\sync}{\texttt{sync()}\xspace}

\newcommand{\seqone}{\texttt{seq-1}\xspace}
\newcommand{\seqtwo}{\texttt{seq-2}\xspace}
\newcommand{\seqthree}{\texttt{seq-3}\xspace}
\newcommand{\dseqthree}{\texttt{seq-3-data}\xspace}
\newcommand{\mseqthree}{\texttt{seq-3-metadata}\xspace}
\newcommand{\nseqthree}{\texttt{seq-3-nested}\xspace}

\newcommand{\xfstests}{\texttt{xfstests}\xspace}
\newcommand{\dmlog}{\texttt{dm-log-writes}\xspace}

\newcommand{\sysioctl}{\texttt{ioctl}\xspace}

\newcommand{\apname}{$B^3$\xspace}
\newcommand{\usysname}{{\small\textsc{Ace}}\xspace}
\newcommand{\lsysname}{{\small\textsc{CrashMonkey}}\xspace}

\newcommand{\fsck}{\texttt{fsck}\xspace}
\newcommand{\lsys}{\lsysname}
\newcommand{\usys}{\usysname}

\newcommand{\twotools}{\lsys and \usys{}\xspace}

%% file: abstract.tex
\begin{abstract}
We present a new approach to testing file-system crash consistency:
\emph{bounded black-box crash testing} (\apname). \apname tests the
file system in a black-box manner using workloads of file-system
operations. Since the space of possible workloads is infinite, \apname
bounds this space based on parameters such as the number of
file-system operations or which operations to include, and
exhaustively generates workloads within this bounded space. Each
workload is tested on the target file system by \emph{simulating}
power-loss crashes while the workload is being executed, and checking
if the file system recovers to a correct state after each
crash. \apname builds upon insights derived from our study of
crash-consistency bugs reported in Linux file systems in the last five
years. We observed that most reported bugs can be reproduced using
small workloads of three or fewer file-system operations on a
newly-created file system, and that all reported bugs result from
crashes after \sysfsync related system calls.  We build two tools,
\twotools, to demonstrate the effectiveness of this approach. Our
tools are able to find 24 out of the 26 crash-consistency bugs
reported in the last five years. Our tools also revealed 10 \emph{new}
crash-consistency bugs in widely-used, mature Linux file systems,
seven of which existed in the kernel since 2014. Our tools also
found a crash-consistency bug in a verified file system, FSCQ. 
The new bugs result
in severe consequences like broken rename atomicity and loss of
persisted files.
\end{abstract}

%% file: intro.tex
\section{Introduction}
\label{sec-intro}

%Why is it important to test crash consistency?
A file system is \emph{crash consistent} if it always recovers to a
correct state after a crash due to a power loss or a kernel panic. The
file-system state is correct if the file system's internal data
structures are consistent, and files that were persisted before the
crash are not lost or corrupted. When developers added delayed
allocation to the ext4 file system~\cite{mathur2007new} in 2009,
they introduced a crash-consistency bug that led to wide-spread data
loss~\cite{URLext4dataloss}.
%An example is the bug in the 
%ext4 file system~\cite{AvantikaEtAl07-ext4}
%(introduced when adding the delayed allocation feature in 2009) that
%caused wide-spread data loss before it was
%fixed~\cite{URLext4dataloss}. 
Given the potential consequences of crash-consistency bugs and the
fact that even professionally-managed datacenters occasionally suffer
from power losses~\cite{ power-amazon, power-canada, power-dreamhost,
  power-fire, power-internap, power-london}, it is important to ensure
that file systems are crash consistent.

% What is lacking today?
Unfortunately, there is little to no crash-consistency testing today
for widely-used Linux file systems such as ext4,
xfs~\cite{sweeneyetal96-xfs}, btrfs~\cite{rodeh2013btrfs}, and
F2FS~\cite{lee2015f2fs}. The current practice in the Linux file-system
community is to not do any \emph{proactive} crash-consistency
testing. If a user reports a crash-consistency bug, the file-system
developers will then \emph{reactively} write a test to capture that
bug. Linux file-system developers use \xfstests~\cite{xfstests}, an
ad-hoc collection of correctness tests, to perform regression
testing. \xfstests contains a total of 482 correctness tests that are
applicable to all POSIX file systems. Of these 482 tests, only 26
(5\%) are crash-consistency tests. Thus, file-system developers have
no easy way of systematically testing the crash consistency of their
file systems.

%Our solution
This paper introduces a new approach to testing file-system crash
consistency: \emph{bounded black-box crash testing} (\apname).
\apname is a black-box testing approach: no file-system code is
modified. \apname works by exhaustively generating workloads within a
bounded space, \emph{simulating} a crash after persistence operations
like \sysfsync in the workload, and finally testing whether the file
system recovers correctly from the crash. We implement the \apname
approach by building two tools, \twotools. Our tools are able to find
24 out of the 26 crash-consistency bugs reported in the last five
years, across seven kernel versions and three file systems.
Furthermore, the systematic nature of \apname allows our tools to find
\emph{new} bugs: \twotools find 10 bugs in widely-used Linux file
systems which lead to severe consequences such as \texttt{rename()}
not being atomic and files disappearing after \sysfsync and a data loss bug in
the FSCQ verified file system~\cite{fscq}. We have
reported all new bugs; developers have submitted patches for five, and
are working to fix the rest. 
% Talk about the study
We formulated \apname based on our study of all 26 crash-consistency
bugs in ext4, xfs, btrfs, and F2FS reported in the last five years
(\sref{sec-study}). Our study provided key insights that made \apname
feasible: most reported bugs involved a small number of file-system
operations on a new file system, with a crash right after a
\emph{persistence point} (a call to \sysfsync, \sysfdsync, or
\texttt{sync} that flushes data to persistent storage). Most bugs
could be found or reproduced simply by systematic testing on a small
space of workloads, with crashes only after persistence points. Note
that without these insights which bound the workload space, \apname is
infeasible: there are infinite workloads that can be run on infinite
file-system images.

Choosing to crash the system only after persistence points is one
of the key decisions that makes \apname tractable.  \apname does not
explore bugs that arise due to crashes in the \emph{middle} of a
file-system operation because file-system guarantees are undefined in
such scenarios. Moreover, \apname cannot reliably assume that the
on-storage file-system state has been modified if there is no
persistence point. Crashing only after persistence points bounds the
work to be done to test crash consistency, and also provides clear
correctness criteria: files and directories which were successfully
persisted before the crash must survive the crash and not be
corrupted.% work on this a bit more

\apname \emph{bounds} the space of workloads in several other
ways. First, \apname restricts the number of file-system operations in
the workload, and simulates crashes only after persistence points.
Second, \apname restricts the files and directories that function as
arguments to the file-system operations in the workload. Finally,
\apname restricts the initial state of the system to be a small, new
file system. Together, these bounds greatly reduce the space of
possible workloads, allowing \twotools to exhaustively generate and
test workloads.

%The class of bugs we focus on
%\apname focuses on a specific class of bugs that result from
%a crash after a call to persistence operations like \sysfsync.

%Contribution 2 : CrashMonkey
An approach like \apname is only feasible if
we can \emph{automatically} and \emph{efficiently} check crash
consistency for arbitrary workloads. We built \lsysname, a framework
that simulates crashes during workload execution and tests for
consistency on the recovered file-system image.
%% \if 0
%% When testing consistency,
%% a fine-grained checker is used to validate the data and
%% metadata of peristed files and directories instead of using a file-system checker
%% like \texttt{fsck}~\cite{mckusicketal86-fsck}.
%% \fi
%% \if 0
%% \fsck is avoided
%% because
%% it's run time depends upon the total amount of
%% data in the file system, even if the test workload only touched one or
%% two files. \texttt{fsck} is also insufficient to check for data loss. 
%% For example, a bug which results in the file system
%% returning to an empty state after a crash would be missed by
%% \texttt{fsck}.
%% \fi
\lsysname first profiles a given workload, capturing all the IO
resulting from the workload. It then replays IO requests until a
persistence point to create a new file-system image we term a
\emph{crash state}.  At each persistence point, \lsysname also
captures a snapshot of files and directories which have been
explicitly persisted (and should therefore survive a crash). \lsysname
then mounts the file system in each crash state, allows the file
system to recover, and uses it's own fine-grained checks to validate
if persisted data and metadata are available and correct. Thus,
\lsysname is able to check crash consistency for arbitrary workloads
automatically, without any manual effort from the user. This property
is key to realizing the \apname approach.

%Contribution 3 : ACE
We built the Automatic Crash Explorer (\usysname) to exhaustively
generate workloads given user constraints and file-system
semantics. \usysname first generates a sequence of file-system
operations; \eg a \texttt{link()} followed by a
\texttt{rename()}. Next, \usys fills in the arguments of each
file-system operation. It then exhaustively generates workloads where
each file-system operation can optionally be followed by an \sysfsync,
\texttt{fdatasync()}, or a global \texttt{sync} command. Finally,
\usys adds operations to satisfy any dependencies (\eg a file must
exist before being renamed). Thus, given a set of constraints,
\usysname generates an exhaustive set of workloads, each of which is
tested with \lsysname on the target file system.

% related work and limitations
\apname offers a new point in the spectrum of techniques addressing
file-system crash consistency, alongside verified file
systems~\cite{fscq, tree-verify, verify-push} and model
checking~\cite{fisc, YangEtAl06-Explode}. Unlike these approaches,
\apname targets widely deployed file systems written in low-level
languages, and does not require annotating or modifying file-system
code.

However, \apname is not without limitations as it is not guaranteed to
find all crash-consistency bugs.  Currently, \usys{}'s bounds do not
expose bugs that require a large number of operations or exhaustion of
file-system resources.  While \lsys can test such a workload, \usys
will not be able to automatically generate the workload. Despite these
limitations, we are hopeful that the black-box nature and ease-of-use
of our tools will encourage their adoption in the file-system
community, unlike model checking and verified file systems. We are
encouraged that researchers at Hanyang University are using our tools
to test the crash consistency of their research file system,
BarrierFS~\cite{won2018barrier}.

This paper makes the following contributions:
\begin{compactitem}
\item A detailed analysis of crash-consistency bugs reported across
  three widely-used file systems and seven kernel versions in the last
  five years (\sref{sec-study})
\item The bounded black-box crash testing approach (\sref{sec-ap})
\item The design and implementation of \twotools \footnote{
\url{https://github.com/utsaslab/crashmonkey}} (\sref{sec-tools})
\item Experimental results demonstrating that our tools are able to
  efficiently find existing and new bugs across widely-used Linux file
  systems and verified file systems. (\sref{sec-eval})
\end{compactitem}  

%% file: mot2.tex
\section{Background}
\label{sec-mot}

We first provide some background on file-system crash consistency, why
crash-consistency bugs occur, and why it is important to test
file-system crash consistency.

\vheading{Crash consistency}.  A file system is crash-consistent if a
number of invariants about the file-system state hold after a crash
due to power loss or a kernel panic~\cite{mckusick1999soft,
  ChidambaramPhd15}.  Typically, these invariants include using
resources only after initialization (\eg path-names point to
initialized metadata such as inodes), safely reusing resources after
deletion (\eg two files shouldn't think they both own the same data
block), and atomically performing certain operations such as renaming
a file. Conventionally, crash consistency is only concerned with
internal file-system integrity. A bug that loses previously persisted
data would not be considered a crash-consistency bug as long as the
file system remains internally consistent. In this paper, we widen the
definition to include data loss. Thus, if a file system loses
persisted data or files after a crash, we consider it a
crash-consistency bug. The Linux file-system developers agree with
this wider definition of crash consistency~\cite{data-guarantee1,
  data-guarantee2}. However, it is important to note that data or
metadata that has not been \emph{explicitly persisted} does not fall
under our definition; file systems are allowed to lose such data in
case of power loss. Finally, there is an important difference between
crash-consistency bugs and file-system correctness bugs:
crash-consistency bugs \emph{do not} lead to incorrect behavior if no
crash occurs.

\vheading{Why crash-consistency bugs occur}. The root of crash
consistency bugs is the fact that most file-system operations
\emph{only modify in-memory state}. For example, when a user creates a
file, the new file exists only in memory until it is explicitly
persisted via the \sysfsync call or by a background thread which periodically writes
out dirty in-memory data and metadata.

Modern file systems are complex and keep a
significant number of metadata-related data structures in
memory. For example, btrfs organizes its metadata as B+
trees~\cite{rodeh2013btrfs}. Modifications to these data structures are accumulated in
memory and written to storage either on \sysfsync, or by a background
thread. Developers could make two common types of mistakes while persisting these
in-memory structures, which consequently lead to crash-consistency bugs. 
The first is neglecting to update certain fields of the data structure.
For example, btrfs had a bug where the field in the 
file inode that determined whether it
should be persisted was not updated. As a result, \sysfsync on the file became a
no-op, causing data loss on a crash~\cite{btrfs-generic59}. The second
is improperly ordering data and metadata when
persisting it. For example, when delayed allocation was introduced in
ext4, applications that used rename to atomically update files
lost data since the rename could be persisted before the file's new
data~\cite{URLext4dataloss}. Despite the fact that the errors that cause
crash-consistency bugs are very different in these two
cases, the fundamental problem is that some in-memory state that is 
required to recover correctly is not written to disk.

\vheading{POSIX and file-system guarantees}. Nominally, Linux file
systems implement the POSIX API, providing guarantees as laid out in
the POSIX standard~\cite{posix7}. Unfortunately, POSIX is extremely
vague. For example, under POSIX it is legal for \sysfsync to
\emph{not} make data durable~\cite{sys-fsync}. Mac OSX takes advantage
of this legality, and requires users to employ
\texttt{fcntl(F\_FULLFSYNC)} to make data durable~\cite{mac-fsync}. As
a result, file systems often offer guarantees above and beyond what is
required by POSIX. For example, on ext4, persisting a new file will
also persist its directory entry. Unfortunately, these guarantees vary
across different file systems, so we contacted the developers of each
file system to ensure we are testing the guarantees that they seek to
provide.

\input{tbl-example}

\vheading{Example of a crash-consistency
  bug}. Figure~\ref{fig-example} shows a crash-consistency bug in
btrfs that causes the file system to become un-mountable (unavailable)
after the crash. Resolving the bug requires file-system repair using
\texttt{btrfs-check}; for lay users, this requires guidance of the
developers~\cite{btrfs-check}. This bug occurs on btrfs because the
unlink affects two different data structures which become out of sync
if there is a crash. On recovery, btrfs tries to unlink \texttt{bar}
twice, producing an error.

\vheading{Why testing crash consistency is important}. File-system
researchers are developing new crash-consistency
techniques~\cite{ChidambaramEtAl13-OptFS, Chidambaram+12-NoFS,
  PillaiEtAl17-CCFS} and designing new file systems that increase
performance~\cite{wafl-fast17, ext4-shingled, betrfs,
  decoupling-fast16, HpTransProc-fast18, FStream-fast18,
  DBLP:conf/sosp/BhatECKZ17, HuEtAl18-TxFS}. Meanwhile, Linux file
systems such as btrfs include a number of optimizations that affect
the ordering of IO requests, and hence, crash consistency. However,
crash consistency is subtle and hard to get right, and a mistake could
lead to silent data corruption and data loss. Thus, changes affecting
crash consistency should be carefully tested.

\vheading{State of crash-consistency testing
  today}. \xfstests~\cite{xfstests} is a regression test suite to
check file-system correctness, with a small proportion (5\%) of
crash-consistency tests. These tests are aimed at avoiding the
recurrence of the same bug over time, but do not generalize to
identifying variants of the bug.  Additionally, each of these test
cases requires the developer to write a checker describing the correct
behavior of the file system after a crash. Given the infinite space of
workloads, it is extremely hard to handcraft workloads that could
reveal bugs. These factors make \xfstests insufficient to identify
\emph{new} crash-consistency bugs.

%% file: tbl-example.tex
%% \begin{table}[!t]
%%   \small
%%   \centering
%%   \ra{1.3}
%%     \begin{tabular}{@{}c@{}}
%%       \toprule[1.2pt]
%%       \begin{verbatim}
%%         hello
%%       \end{verbatim}\\
%%     \bottomrule[1.2pt]
%%     \end{tabular}
%%   \label{tbl-example}
%% \end{table}

\begin{figure}
\begin{Verbatim}[fontsize=\small]
1 create foo
2 link foo bar  
3 sync
4 unlink bar
5 create bar
6 fsync bar
7 CRASH!
\end{Verbatim}
\vspace{-0.2in}
\mycaption{Example crash-consistency bug}{The figure shows the workload
  to expose a crash-consistency bug that was reported in the btrfs file system in
  Feb 2018~\cite{btrfs-generic-link-unlink}. The bug causes the file
  system to become un-mountable. }
  %Resolving the bug requires repairing
 % the file system with \texttt{btrfs-check} with developer advisement
%  (the btrfs developers do not suggest the general public use
  %btrfs-check on their own~\cite{btrfs-check}).}
\label{fig-example}
\vspace{-10pt}
\end{figure}

%% file: study.tex
\section{Studying Crash-Consistency Bugs}
\label{sec-study}

We present an analysis of 26 unique crash-consistency bugs reported by
users over the last five years on widely-used Linux file
systems~\cite{cm-reproduced}. We find these bugs either by examining
mailing list messages or looking at the crash-consistency tests in the
\xfstests regression test suite. Few of the crash-consistency tests in
\xfstests link to the bugs that resulted in the test being written.

Due to the nature of crash-consistency bugs (all in-memory information
is lost upon crash), it is hard to tie them to a specific workload.
As a result, the number of reported bugs is low. We believe there are
many crash-consistency bugs that go unreported in the wild.

\input{tbl-bugs}
\input{tbl-reproduced-bugs}

We analyze the bugs based on consequence, kernel version, file system,
and the number of file-system operations required to reproduce
them. There are 26 unique bugs spread across ext4, F2FS, and
btrfs. Each unique bug requires a unique set of file-system operations
to reproduce. Two bugs occur on two file systems (F2FS and
ext4, F2FS and btrfs), leading to a total of 28 bugs.

Table~\ref{tbl-bugs} presents some statistics about the
crash-consistency bugs. The table presents the kernel version in which
the bug was reported. If the bug report did not include a version, it
presents the latest kernel version in which \apname could reproduce
the bug (the two bugs that \apname could not reproduce appear in
kernel 3.13). The bugs have severe consequences, ranging from
file-system corruption to the file system becoming un-mountable.  The
four most common file-system operations involved in crash-consistency
bugs were \syswrite, \syslink, \sysunlink, and \sysrename. Most
reported bugs resulted from either reusing filenames in multiple
file-system operations or write operations to overlapping file
regions. Most reported bugs could be reproduced with three or fewer
file-system operations.
% though two bugs additionally required
%either a specific file-system image or a special command (\texttt{dropcaches})
%to reproduce. 
%The highest  bugs are found in btrfs, which we attribute to its increasing
%complexity over time. This trend is reflected in crash-consistency bugs reported in
% recent kernel versions.

\vheading{Examples}. Table~\ref{tbl-reproduced-bugs} showcases a few
of the crash-consistency bugs. Bug \#1~\cite{btrfs-generic34} involves creating two files in
a directory and persisting only one of them. btrfs log recovery
incorrectly counts the directory size, making the directory
un-removable thereafter. Bug \#2~\cite{btrfs-generic56} involves creating a hard link to an
already existing file. A crash results in btrfs recovering the file
with a size 0, thereby making its data inaccessible.  A similar bug
(\#5~\cite{ext4-direct-write}) manifests in ext4 in the direct write path, where the write
succeeds and blocks are allocated, but the file size is incorrectly
updated to be zero, leading to data loss.

\vheading{Complexity leads to bugs}.  The ext4 file system has
undergone more than 15 years of development, and, as a result, has
only two bugs. The btrfs and F2FS file systems are more recent: btrfs
was introduced in 2007, while F2FS was introduced in 2012. In
particular, btrfs is an extremely complex file system that provides
features such as snapshots, cloning, out-of-band deduplication, and
compression. btrfs maintains its metadata (such as inodes and bitmaps)
in the form of various copy-on-write B+ trees. This makes achieving
crash consistency tricky, as the updates have to be propagated to
several trees.  Thus, it is not surprising that most reported
crash-consistency bugs occurred in btrfs. As file systems become more
complex in the future, we expect to see a corresponding increase in
crash-consistency bugs.

\vheading{Crash-consistency bugs are hard to find}. Despite the fact
that the file systems we examined were widely used, some bugs have
remained hidden in them for years. For example, btrfs had a
crash-consistency bug that was only discovered \emph{seven} years
after it was introduced. The bug was caused by incorrectly processing
a hard link in btrfs's data structures. When a hard link is added, the
directory entry is added to one data structure, while the inode is
added to another data structure. When a crash occurred, only one of
these data structures would be correctly recovered, resulting in the
directory containing the hard link becoming
un-removable~\cite{btrfs-generic-unremovable}. This bug was present
since the log tree was added in 2008; however, the bug was only
discovered in 2015.

\vheading{Systematic testing is required}. Once the hard link bug in
btrfs was discovered, the btrfs developers quickly fixed it. However,
they only fixed one code path that could lead to the bug.  The same
bug could be triggered in another code path, a fact that was only
discovered \emph{four months} after the original bug was
reported. While the original bug workload required creating hard links
and calling \sysfsync on the original file and parent directory, this one
required calling \fsync{} on a sibling in the directory where the hard
link was created~\cite{btrfs-generic104}. Systematic testing of the
file system would have revealed that the bug could be triggered via an
alternate code path.

\vheading{Small workloads can reveal bugs on an empty file
  system}. Most of the reported bugs do not require a special
file-system image or a large number of file-system operations to
reproduce. 24 out of the 26 reported bugs require three or fewer core
file-system operations to reproduce on an empty file system. This
count is low because
we do not count \emph{dependent} operations: for example, a file has
to exist before being renamed and a directory has to exist before a
file can be created inside it. Such dependent operations can be
\emph{inferred} given the core file-system operations. Of the
remaining two bugs, one required a special command
(\texttt{dropcaches}) to be run during the workload for the bug to
manifest. The other bug required specific setup: 3000 hard
links had to already exist (forcing an external reflink) for the bug
to manifest.

\vheading{Reported bugs involve a crash after persistence}.  All
reported bugs involved a crash right after a persistence point: a call
to \sysfsync, \sysfdsync, or the global \texttt{sync} command. These
commands are important because file-system operations only modify
in-memory metadata and data by default. Only persistence points
reliably change the file-system state on storage. Therefore, unless a
file or directory has been persisted, it cannot be expected to survive
a crash. While crashes could technically occur at any point, a user
cannot complain if a file that has not been persisted goes missing
after a crash. Thus, every crash-consistency bug involves
\emph{persisted data or metadata} that is affected by the bug after a
crash, and a workload that does not have a persistence point cannot
lead to a reproducible crash-consistency bug. This also points to an
effective way to find crash-consistency bugs: perform a sequence of
file-system operations, change on-storage file-system state with
\sysfsync or similar calls, crash, and then check files and
directories that were previously persisted.

%% file: tbl-bugs.tex
\begin{table}[!t]
  \small
  \centering
  \ra{1.3}
  \begin{tabular}{@{}p{92pt}r@{}}
%\begin{tabularx}{50pt}{@{}lrp{2cm}@{}}    
    \toprule[1.2pt]
    Consequence & \# bugs \\
    \midrule
    Corruption &  19 \\
    Data Inconsistency & 6 \\
    Un-mountable file system & 3 \\
    \midrule
    Total & 28 \\
    \bottomrule[1.2pt]
   %\end{tabularx}
\end{tabular}
\hfill
    \begin{tabular}{@{}lr@{}}
    \toprule[1.2pt]
    Kernel Version & \# bugs \\
    \midrule
    3.12 & 3 \\
    3.13 & 9 \\
    3.16 & 1 \\
    4.1.1 & 2 \\
    4.4 & 9 \\
    4.15 & 3 \\
    4.16 (latest) & 1 \\
    \midrule
    Total & 28 \\
    %\bottomrule[1.2pt]
    \end{tabular}
    
    \begin{tabular}{@{}lr@{}}
    \toprule[1.2pt]
    File System & \# bugs \\
    \midrule
    ext4 & 2 \\
    F2FS & 2 \\
    btrfs & 24 \\
    \midrule
    Total & 28 \\
    \bottomrule[1.2pt]
    \end{tabular}
    \hfill
  \begin{tabular}{@{}rr@{}}
    \toprule[1.2pt]
    \# of ops required & \# bugs \\
    \midrule
    1 & 3 \\
    2 & 14 \\
    3 & 9 \\
    \midrule
    Total & 26 \\
    \bottomrule[1.2pt]
  \end{tabular}

  \mycaption{Analyzing crash-consistency bugs}{The tables break down
    the 26 unique crash-consistency bugs reported over the last five
    years (since 2013) by different criteria. Two bugs were reported
    on two different file systems, leading to a total of 28 bugs.}
  \label{tbl-bugs}
\end{table}

%% file: tbl-reproduced-bugs.tex
\begin{table*}[!t]
  \small
  \centering
  \ra{1.3}
  \begin{tabular}{@{}rllrr@{}}
    \toprule[1.2pt]
    Bug \# & File System & Consequence & \# of ops & ops involved (excluding persistence operations)\\
    \midrule
    1 & btrfs & Directory un-removable & 2 & \texttt {{creat(A/x)}}, \texttt{creat(A/y)}\xspace \\
    2 & btrfs & Persisted data lost & 2 & \texttt {pwrite(x)}, \texttt{link(x,y)}\xspace\\
    3 & btrfs & Directory un-removable & 3 & \texttt{ link(x,A/x)}, \texttt{link(x,A/y)}, \texttt{unlink(A/y)}\xspace\\
    4 & F2FS & Persisted file disappears & 3 & \texttt{ pwrite(x)}, \texttt{rename(x,y)}, \texttt{pwrite(x)}\xspace\ \\
    5 & ext4 & Persisted data  lost & 2 & \texttt{ pwrite(x)}, \texttt{direct\_write(x)}\xspace \\
    \bottomrule[1.2pt]
  \end{tabular}
  \vspace{-5pt}  
  \mycaption{Examples of crash-consistency bugs}{The table shows some
    of the crash-consistency bugs reported in the last five years. The
    bugs have severe consequences, ranging from losing user data to
    making directories un-removable.}
  \label{tbl-reproduced-bugs}
\vspace{-0.1in}  
\end{table*}

%% file: ap4.tex
\section{\apname: Bounded Black-Box Crash Testing}
\label{sec-ap}

% B^3 is something that specifies bounds for workloads, that the sytem
% should exhaustively generate workloads given the bounds, that a
% crash test should be run after every persistence point, and that
% only the set of files that were affected by the persistence
% operation should be checked in the resulting recovered crash state.

Based on the insights from our study of crash-consistency bugs, we
introduce a new approach to testing file-system crash consistency:
\emph{Bounded Black-Box crash testing} (\apname). \apname is a
black-box testing approach built upon the insight
that most reported crash-consistency bugs can be found by
systematically testing small sequences of file-system operations on a
new file system. \apname exercises the file system through its
system-call API, and observes the file-system behavior via read and
write IO. As a result, \apname does not require annotating or
modifying file-system source code.

\subsection{Overview}

% B^3 should not generate anything, it is just an approach. Use
% something like 'focuses' or 'centers on' instead?

% B^3 is a specification on how to exhaustively generate and test file
% system workloads?
\apname generates sequences of file-system operations, called
\emph{workloads}. Since the space of possible workloads is infinite,
\apname \emph{bounds} the space of workloads using insights from the
study. Within the determined bounds, \apname exhaustively generates
and tests all possible workloads. Each workload is tested by
\emph{simulating} a crash after each persistence point, and checking
if the file system recovers to a correct state. \apname performs
fine-grained correctness checks on the recovered file-system state;
only files and directories that were explicitly persisted are
checked. \apname checks for both data and metadata (size, link count,
and block count) consistency for files and directories.

 \vheading{Crash points}. The main insight from the study that makes
 an approach like \apname feasible is the choice of crash points; a
 crash is simulated \emph{only} after each persistence point in the
 workload instead of in the middle of file-system operations. This
 design choice was motivated by two factors. First, file-system
 guarantees are undefined if a crash occurs in the middle of a
 file-system operation; only files and directories that were
 previously successfully persisted need to survive the
 crash. File-system developers are overloaded, and bugs involving data
 or metadata that has not been explicitly persisted is given low
 priority (and sometimes not acknowledged as a bug). Second, if we
 crash in the middle of an operation, there are a number of correct
 states the file system could recover to. If a file-system operation
 translates to \emph{n} block IO requests, there could be \emph{$2^n$}
 different on-disk crash states if we crashed anywhere during the
 operation. Restricting crashes to occur after persistence points
 bounds this space linearly in the number of operations comprising the
 workload. The small set of crash points and correct states makes
 automated testing easier. Our choice of crash points naturally leads
 to bugs where persisted data and metadata is corrupted or missing and
 file-system developers are strongly motivated to fix such bugs.

\subsection{Bounds used by \apname}

Based on our study of crash-consistency bugs, \apname bounds the space
of possible workloads in several ways:
\begin{compactenum}

\item \vheading{Number of operations}. \apname bounds the number of
  file-system operations (termed the \emph{sequence length}) in the
  workload. A \texttt{seq-X} workload has $X$ core file-system
  operations in it, not counting dependent operations such as creating
  a file before renaming it.

\item \vheading{Files and directories in workload}. We observe that in
  the reported bugs, errors result from the \emph{reuse} of a small set
  of files for metadata operations. Thus, \apname restricts workloads
  to use few files per directory, and a low directory depth.
  This restriction automatically reduces the inputs for metadata-related
  operations such as \sysrename. 
  
\item \vheading{Data operations}. The study also indicated that bugs 
related to data inconsistency mainly occur due to writes to 
\emph{overlapping} file ranges. In most cases,
  the bugs are not dependent on the exact offset and length used in the
  writes, but on the interaction between the overlapping regions
  from writes. The study indicates that a broad classification of writes such as
  appends to the end of a file, overwrites to overlapping regions of file, \etc 
  is sufficient to find crash-consistency bugs.
  
\item \vheading{Initial file-system state}. Most of the bugs analyzed
  in the study did not require a specific initial file-system state (or
  a large file system) to be revealed. Moreover, most of the
  studied bugs could be reproduced starting from the
  \emph{same, small} file-system image. Therefore, \apname can test all workloads
  starting from the same initial file-system state.

\end{compactenum}  
%\vspace{-1pt}

\subsection{Fine-grained correctness checking}
\label{sec-ap-check}
\apname uses fine-grained correctness checks to validate the data and
metadata of persisted files and directories in each crash state. Since
\fsck is both time-consuming to run and can miss data loss/corruption bugs,
it is not a suitable checker for \apname.

\subsection{Limitations}
\label{sec-limit}

The \apname approach has a number of limitations:
\begin{compactenum}
\item \apname does not make any guarantees about finding \emph{all}
  crash-consistency bugs. It is sound but incomplete. However, because \apname
  tests exhaustively, if the workload that triggers the bug falls
  within the constrained workload space, \apname will find it. Therefore, the
  effectiveness of \apname depends upon the bounds chosen and the
  number of workloads tested.
  
\item  \apname focuses on a specific class of bugs. It does not
  simulate a crash in the middle of a file-system operation and it does not
  re-order IO requests to create different crash states. The implicit
  assumption is that the core crash-consistency mechanism, such as
  journaling~\cite{PrabhakaranEtAl05-Usenix} or
  copy-on-write~\cite{HitzEtAl94-Netapp, RosenblumOusterhout92-LFS},
  is working correctly. Instead, we assume that it is the rest of the file
  system that has bugs. The crash-consistency bug study indicates this
  assumption is reasonable.
  
\item \apname focuses on workloads where files and directories are
  explicitly persisted. If we created a file, waited one hour, then
  crashed, and found that the file was gone after the file-system
  recovered, this would also be a crash-consistency bug. However,
  \apname does not explore such workloads as they take a significant
  amount of time to run and are not easily reproduced in a
  deterministic fashion.
  
\item Due to its black-box nature, \apname cannot pinpoint the
  exact lines of code that result in the observed bug. Once a bug has
  been revealed by \apname, finding the root cause requires further
  investigation. However, \apname aids in investigating the root cause of the
  bug since it
  provides a way to reproduce the bug in
  a deterministic fashion.
\end{compactenum}

Despite its shortcomings, we believe \apname is a useful addition 
to the arsenal of techniques for testing file-system crash consistency.
The true strengths of \apname lie in its
systematic nature and the fact that it does not require any changes to
existing systems. Therefore, it is ideal for complex and widely-used
file systems written in low-level languages like C, where stronger
approaches like verification cannot be easily used.

%% file: tools3.tex
% TODO(ashlie): Define crash point better in earlier sections.
\section{CrashMonkey and Ace}
\label{sec-tools}

We realize the \apname approach by building two tools, \twotools.
As shown in Figure~\ref{fig-arch}, \lsysname is responsible for simulating crashes at
different points of a given workload and testing if the file system
recovers correctly after each simulated crash, while the Automatic Crash Explorer
(\usysname) is responsible for exhaustively generating workloads in a bounded space.

\input{fig-arch}

\subsection{CrashMonkey}
\lsys uses record-and-replay techniques to \emph{simulate} a crash in the middle
of the workload and test if the file system recovers to a correct state after
the crash. For maximum portability, \lsys treats the file system as a
black box, only requiring that the file system implement the POSIX API.

\vheading{Overview}. \lsys operates in three phases as shown in
Figure~\ref{fig-cm}. In the first phase, \lsys profiles the workload
by collecting information about all file-system operations and IO
requests made during the workload. The second phase replays IO
requests until a persistence point to create a \emph{crash state}. The
crash state represents the state of storage if the system had crashed
after a persistence operation completed. \lsys then mounts the file
system in the crash state and allows the file system to perform
recovery. At each persistence point, \lsys also captures a reference
file-system image, termed the \emph{oracle}, by safely unmounting it
so the file system completes any pending operations or checkpointing.
The oracle represents the expected state of the file system after a
crash. In the absence of bugs, persisted files should be the same in
the oracle and the crash state after recovery. In the third phase,
\lsys{}'s AutoChecker tests for correctness by comparing the persisted
files and directories in the oracle with the crash state after
recovery.

\input {fig-cm}

\lsys is implemented as two kernel modules and a set of user-space
utilities. The kernel modules consist of 1300 lines of C code which
can be compiled and inserted into the kernel at run time, thus
avoiding the need for long kernel re-compilations. The user-space
utilities consist of 4800 lines of C++ code. \lsysname{}'s separation
into kernel modules and user-space utilities allows rapid porting to a
different kernel version; only the kernel modules need to be ported to
the target kernel. This allowed us to port \lsysname to seven kernels to
reproduce the bugs studied in \sref{sec-study}.

\input{tbl-ace-bounds}

\vheading{Profiling workloads}. \lsys profiles workloads at two levels
of the storage stack: it records block IO requests, and it records
system calls. It uses two kernel modules to record block IO requests
and create crash states and oracles.

The first kernel module records all IO requests generated by the
workload using a wrapper block device on which the target file system
is mounted. The wrapper device records both data and metadata for IO
requests (such as sector number, IO size, and flags). Each persistence
point in the workload causes a special \emph{checkpoint} request to be
inserted into the stream of IO requests recorded.  The checkpoint is
simply an empty block IO request with a special flag, to correlate the
completion of a persistence operation with the low-level block IO
stream.  All the data recorded by the wrapper device is communicated
to the user-space utilities via \sysioctl calls.

The second kernel module in \lsys is an in-memory, copy-on-write block
device that facilitates snapshots. \lsys creates a snapshot of the
file system before the profiling phase begins, which represents the
base disk image.  \lsys provides fast, writable snapshots by replaying
the IO recorded during profiling on top of the base disk image to
generate a crash state.  Snapshots are also saved at each persistence
point in the workload to create oracles. Furthermore, since the
snapshots are copy-on-write, resetting a snapshot to the base image
simply means dropping the modified data blocks, making it efficient.

\lsys also records all \sysopen, \sysclose, \sysfsync, \sysfdsync,
\sysrename, \sync, and \msync calls in the workload so that when the
workload does a persistence operation such as \texttt{fsync(fd)},
\lsys is able to correlate \texttt{fd} with a file that was opened
earlier. This allows \lsys to track the set of files and directories
that were explicitly persisted at any point in the workload. This
information is used by \lsys{}'s AutoChecker to ensure that only files
and directories explicitly persisted at a given point in the workload
are compared. \lsys uses its own set of functions that wrap system
calls which manipulate files to record the required information.

\vheading{Constructing crash states}. To create a crash state, \lsys
starts from the initial state of the file system (before the workload
was run), and uses a utility similar to \sysdd to replay all recorded
IO requests from the start of the workload until the next checkpoint
in the IO stream. The resultant crash state represents the state of
the storage just after the persistence-related call completed on the
storage device. Since the IO stream replay ends directly after the
next persistence point in the stream, the generated crash point
represents a file-system state that is considered uncleanly
unmounted. Therefore, when the file system is mounted again, the
kernel may run file-system specific recovery code.

\vheading{Automatically testing correctness}. \lsys{}'s AutoChecker is
able to test for correctness automatically because it has three key
pieces of information: it knows which files were persisted, it has the
correct data and metadata of those files in the oracle, and it has the
actual data and metadata of the corresponding files in the crash 
state after recovery. Testing correctness is a
simple matter of comparing data and metadata of persisted files in
the oracle and the crash state.

\lsys avoids using \fsck because its runtime is proportional to the
amount of data in the file system (not the amount of data changed) and
it does not detect the loss or corruption of user data.  Instead, when
a crash state is re-mounted, \lsys allows the file system to run its
recovery mechanism, like journal replay, which is usually more
lightweight than \fsck. \fsck is run only if the recovered file system
is un-mountable.  To check consistency, \lsys uses its own \emph{read}
and \emph{write} checks after recovery.  The read checks used by \lsys
confirm that persisted files and directories are accurately
recovered. The write checks test if a bug makes it impossible to
modify files or directories. For example, a btrfs bug made a directory
un-removable due to a stale file handle~\cite{btrfs-generic34}.

Since each file system has slightly different consistency guarantees,
we reached out to developers of each file system we tested, to
understand the guarantees provided by that file system. In some cases,
our conversations prompted the developers to explicitly write down the
persistence guarantees of their file systems for the first
time~\cite{guarantee}. During this process, we confirmed that most
file systems such as ext4 and btrfs implement a stronger set of
guarantees than the POSIX standard.  For example, while POSIX requires
an \sysfsync on both a newly created file and its parent directory to
ensure the file is present after a crash, many Linux file systems do
not require the \sysfsync of the parent directory. Based on the
response from developers, we report bugs that violate the guarantees
each file system aims to provide.
%We intend to encode such filesystem specific requirements in AutoTester in the future.

%The resulting automatic tests in \lsys are file-system specific
%and reflect the guarantees revealed by the developers.
%\am{did we actually make file system specific tests or did we just ignore
%results if we knew it didn't reflect file system guarantees?}
%\jm{The last sentence does not sound right to me. I believe we did not perform 
%^such filesystem specific tests. Infact of the two cases discussed above, 
%AutoTester always requires that the second one hold - fsync of the file persist it, 
%needless of an explicit fsync on parent directory. We currently have no rules set, 
%that would relax this criteria for a strictly-POSIX filesystem.}

\input{fig-ace}

\subsection{Automatic Crash Explorer (Ace)}

\usys exhaustively generates workloads satisfying the given bounds.
\usys has two components, the workload synthesizer and the adapter for
\lsys.

\vheading{Workload synthesizer}. The workload synthesizer exhaustively 
generates workloads within the state space defined by the user specified bounds.
The workloads generated in this stage are represented in a high-level language, 
similar to the one depicted in Figure~\ref{fig-ace}. 

 \vheading{CrashMonkey Adapter}. A custom adapter converts the
 workload generated by the synthesizer into an equivalent C++ test
 file that \lsys can work with. This adapter handles the insertion of
 wrapped file-system operations that \lsys tracks. Additionally, it
 inserts a special function-call at every persistence point, which
 translates to the checkpoint IO.  It is easy to extend \usys to be
 used with other record-and-replay tools like
 \dmlog~\cite{dmlogwrites} by building custom adapters.

Table~\ref{tbl-ace-bounds} shows how we used the insights from the
study to assign specific values for \apname bounds when we run
\usys. Given these bounds, \usys uses a multi-phase process to
generate workloads that are then fed into \lsys.  Figure~\ref{fig-ace}
illustrates the four phases \usys goes through to generate a \seqtwo
workload.

\vheading{Phase 1: Select operations and generate workloads}. \usys
first selects file-system operations for the given sequence length to
make what we term the \emph{skeleton}. By default, file-system operations
can be repeated in the workload. The user may also supply bounds such
as requiring only a subset of file-system operations be used (\eg to
focus testing on new operations). \usys then exhaustively generates
workloads satisfying the given bounds. For example, if the user
specified the \seqtwo workload could only contain six file-system
operations, \usys will generate $6*6 = 36$ skeletons in phase one.

\vheading{Phase 2: Select parameters}. For each skeleton generated in
phase one, \usys then selects the parameters (system-call arguments)
for each file-system operation. By default, \usys uses two files at
the top level and two sub-directories with two files each as arguments
for metadata-related operations. \usys also understands the semantics of
file-system operations and exploits it to eliminate the generation of
\emph{symmetrical} workloads. For example, consider two operations
\texttt{link(foo, bar)}\xspace and \texttt{link(bar, foo)}\xspace.
The idea is to link two files within the same directory, but the order of
file names chosen does not matter. In this example, one of the
workloads would be discarded, thus reducing the total number of
workloads to be tested for the sequence.

For data operations, \usys chooses between whether a write is an
overwrite at the beginning, middle, or end of the file or simply an
append operation. Furthermore, since our study showed that
crash-consistency bugs occur when data operations overlap, \usys tries
to overlap data operations in phase two.

Each skeleton generated in phase one can lead to multiple
workloads (based on different parameters) in phase two. However, at the end of
this phase, each generated workload has a sequence of file-system
operations with all arguments identified.

\vheading{Phase 3: Add persistence points}. \usys optionally adds a
persistence point after each file-system operation in the workload, but
\usys does not require every operation to be followed by a persistence
point. However, \usys ensures that the last operation in a workload is
always followed by a persistence point so that it is not truncated to a 
workload of lower sequence length. The file or directory to be
persisted in each call is selected from the same set of files and
directories used by phase two, and, for each workload generated by
phase two, phase three can generate multiple workloads by adding
persistence points after different sets of file-system operations.

\vheading{Phase 4: Add dependencies}. Finally, \usys satisfies various
dependencies to ensure the workload can execute on a POSIX file
system. For example, a file has to exist before being renamed or
written to. Similarly, directories have to be created if any
operations on their files are involved. Figure~\ref{fig-ace} shows how
\texttt{A}, \texttt{B}, and \texttt{A/foo} are created as dependencies
in the workload. As a result, a \seqtwo workload can have more than
two file-system operations in the final workloads. At the end of this
phase, \usys compiles each workload from the high-level language into
a C++ program that can be passed to \lsys.

\vheading{Implementation}.  \usys consists of 2500 lines of Python
code, and currently supports 14 file-system operations. All bugs
analyzed in our study used one of these 14 file-system operations.
It is straightforward to expand \usys to support more operations.

\vheading{Running Ace with relaxed bounds}. It is easy to relax the
bounds used by \usys to generate more workloads; this comes at the
cost of computational time used to test the extra workloads. Care
should be taken when relaxing the bounds, since the number of
workloads increases at a rapid rate. For example, \usys generates
about 1.5M workloads with three core file-system operations.  Relaxing
the default bound on the set of files and directories to add one
additional nested directory, increases the number of workloads
generated to 3.7M. This simple change results in $2.5\times$ more
workloads. Note that increasing the number file-system operations in
the workload leads to an increase in the number of  phase-1 skeletons
generated, and adding more
files to the argument set increase the number of phase-2 workloads
that can be created. Therefore, the workload space must be carefully
expanded.

\subsection{Testing and Bug Analysis}

\vheading{Testing Strategy}. Given a target file system, we first
exhaustively generate \seqone workloads and test them using \lsys. We
then proceed to \seqtwo , and then \seqthree workloads. By generating
and testing workloads in this order, \lsys only needs to simulate a
crash at one point per workload. For example, even if a \seqtwo
workload has two persistence points, crashing after the first
persistence point would be equivalent to an already-explored \seqone
workload.

\vheading{Analyzing Bug Reports}. One of the challenges with a
black-box approach like \apname is that a single bug could result in
many different workloads failing correctness tests. We present two
cases of multiple test failures in workloads, and how we mitigate
them.

First, workloads in different sequences can fail because of the same
bug.  Our testing strategy is designed to mitigate this: if a bug
causes incorrect behavior with a single file-system operation, it
should be caught by a \seqone workload. Therefore, if we catch a bug
only in a \seqtwo workload, it implies the bug results from the
interaction of the two file-system operations. Ideally, we would run
\seqone, report any bugs, and apply bug-fix patches given by
developers before running \seqtwo. However, for quicker testing, \usys
maintains a database of all previously found bugs which includes the
core file-system operations that produced each bug and the consequence
of the bug. For all new bugs reports generated by \twotools, it first
compares the workload and the consequence with the database of known
bugs. If there is a match, \usys does not report the bug to the user.

%\am{I am not sure what this paragraph is trying to say, could someone
%  revise it and/or explain it to me? It sounds like we try and reduce
%  larger sequences to lower ones so that it is easier to determine the
%cause of a bug, but I'm not sure if that's correct.}

%\jm{We are not trying to reduce the sequence length -  
%we try to look at the skeleton of the workload, which resembles 
%the phase 1 in Figure~\ref{fig-ace}. If this skeleton as well as
%the consequence reported in the bug report matches, then 
%we could essentially be looking at slightly different workloads
%(say with different file names), but leading to the same bug, 
%because of the same cause. }

\input{fig-postproc} Second, similar workloads in the same sequence
could fail correctness tests due to the same bug. For efficient
analysis, we group together bug reports by the consequence (\eg file
missing), and the skeleton (the sequence of core file-system
operations that comprise the workload) that triggered the bug, as
shown in Figure~\ref{fig-postproc}. Using the skeleton instead of the
fully fleshed-out workload allows us to identify similar bugs.  For
example, the bug that causes appended data to be lost will repeat four
times, once with each of the files in our file set. We can group these
bug reports together and only inspect one bug report from each
group. After verifying each bug, we report it to developers.

%% file: fig-arch.tex
\begin{figure}[!t]
  \centering \includegraphics[width=0.45\textwidth]{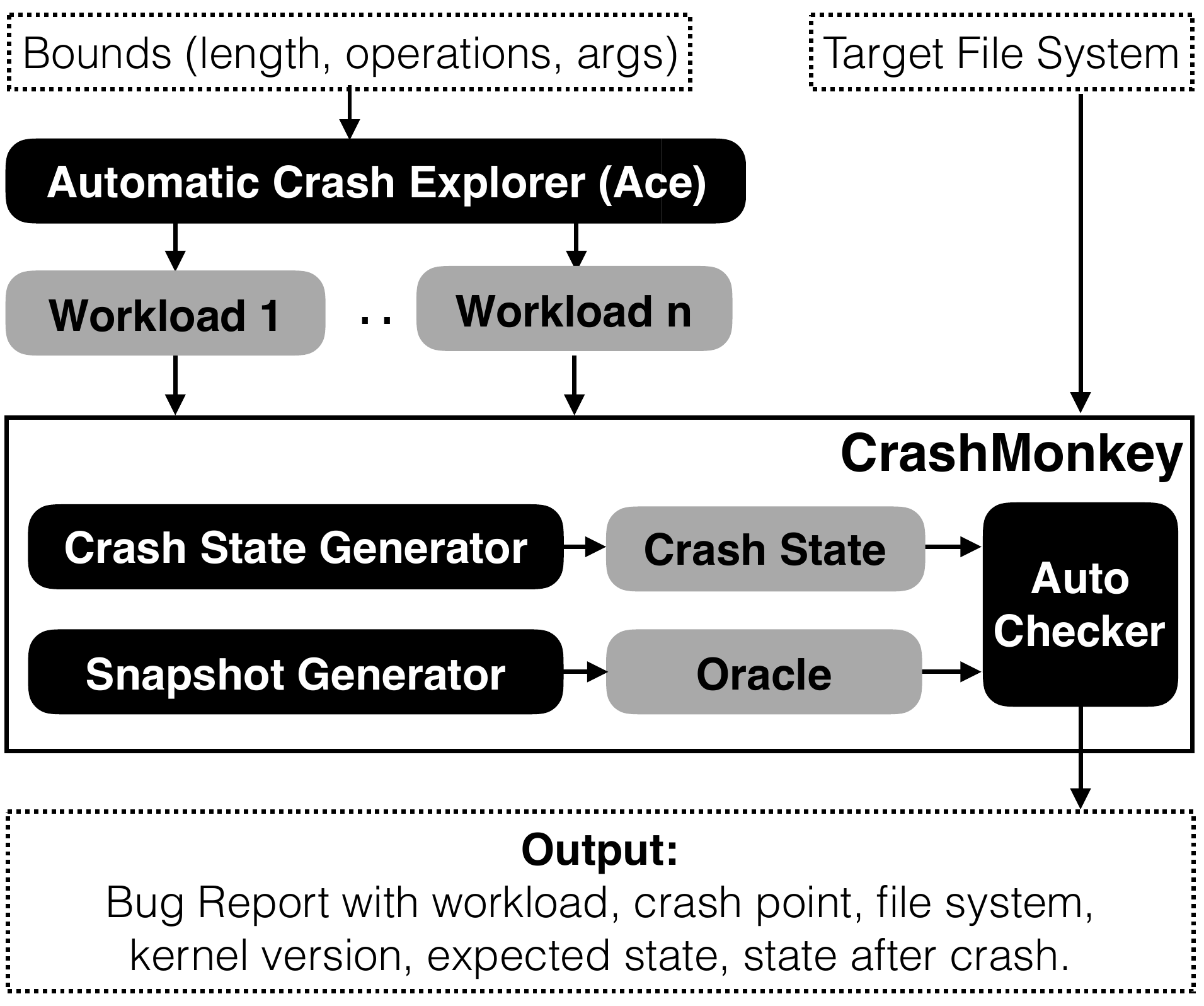}
  \mycaption{System architecture}{Given bounds for exploration,
    \usysname generates a set of workloads. Each workload is then fed
    to \lsysname, which generates a set of crash states and
    corresponding oracles. The AutoChecker compares persisted files in
    each oracle/crash state pair; a mismatch indicates a bug.
    }
  \label{fig-arch}
  \vspace{-10pt}
\end{figure}

%% file: fig-cm.tex
\begin{figure}[!t]
  \centering \includegraphics[width=0.48\textwidth]{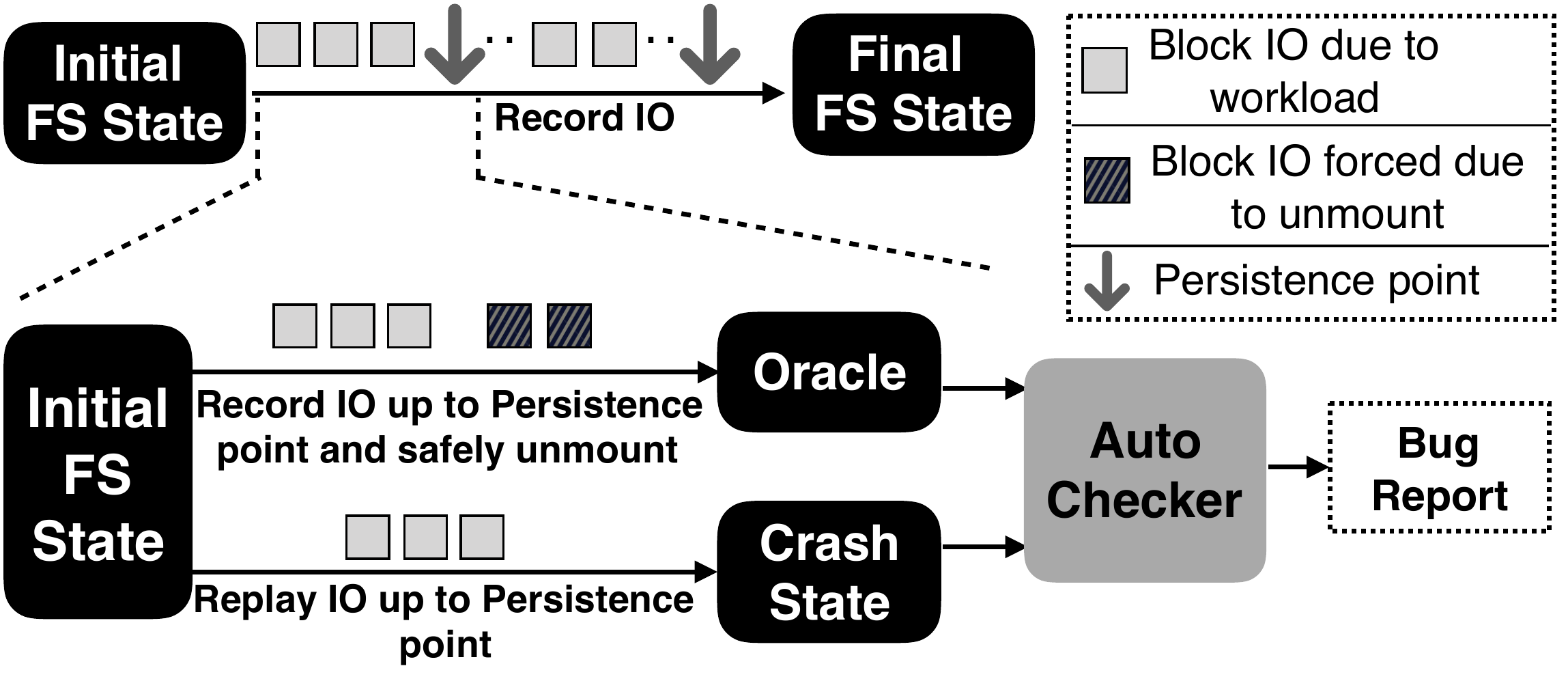}
  \mycaption{\lsys operation}{\lsys first records the block IO requests
  that the workload translates to, capturing reference images
  called oracles after each persistence point.
  \lsys then generates crash states by replaying the recorded IO
  and tests for consistency against the corresponding oracle. }
  \label{fig-cm}
\end{figure}

%% file: tbl-ace-bounds.tex
\begin{table*}[!t]
  \small
  \centering
  \ra{1.3}
  \begin{tabular}{@{}lll@{}}
    \toprule[1.2pt]
    \apname bound  & Insight from the study &  Bound chosen by \usys\\
    \midrule
    Number of operations & Small workloads of 2-3 core operations & Maximum \# of core ops in a workload is \emph{three} \\
    Files and directories & Reuse file and directory names & 2 directories of depth 2, each with 2 unique files \\
    Data operations & Coarse grained, overlapping ranges of writes & Overwrites to start, middle \& end of file, and appends \\
    Initial file-system state & No need of a special initial state or large image & Start with a clean file-system image of size 100MB \\
    \bottomrule[1.2pt]
  \end{tabular}
  \vspace{-5pt}  
  \mycaption{Bounds used by \usys}{The table shows the specific values picked by \usys for each \apname bound.}
  \label{tbl-ace-bounds}
\vspace{-0.1in}  
\end{table*}

%% file: fig-ace.tex
\begin{figure*}[!t]
  \centering \includegraphics[width=0.85\textwidth]{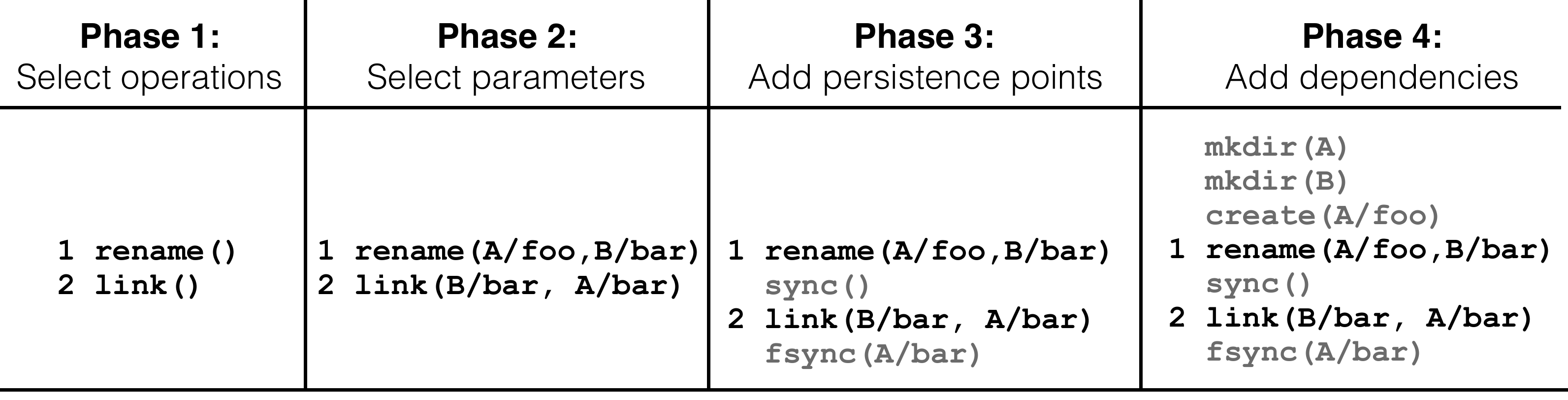}
  \mycaption{Workload generation in \usysname}{The figure shows the
    different phases involved in workload generation in
    \usysname. Given the sequence length, \usysname first selects the
    operations, then selects the parameters for each operation, then
    optionally adds persistence points after each operation, and
    finally satisfies file and directory dependencies for the
    workload. The final workload may have more operations than the
    original sequence length.
  }
  \label{fig-ace}
\end{figure*}

%% file: fig-postproc.tex
\begin{figure}[!t]
  \centering
  \includegraphics[width=0.45\textwidth]{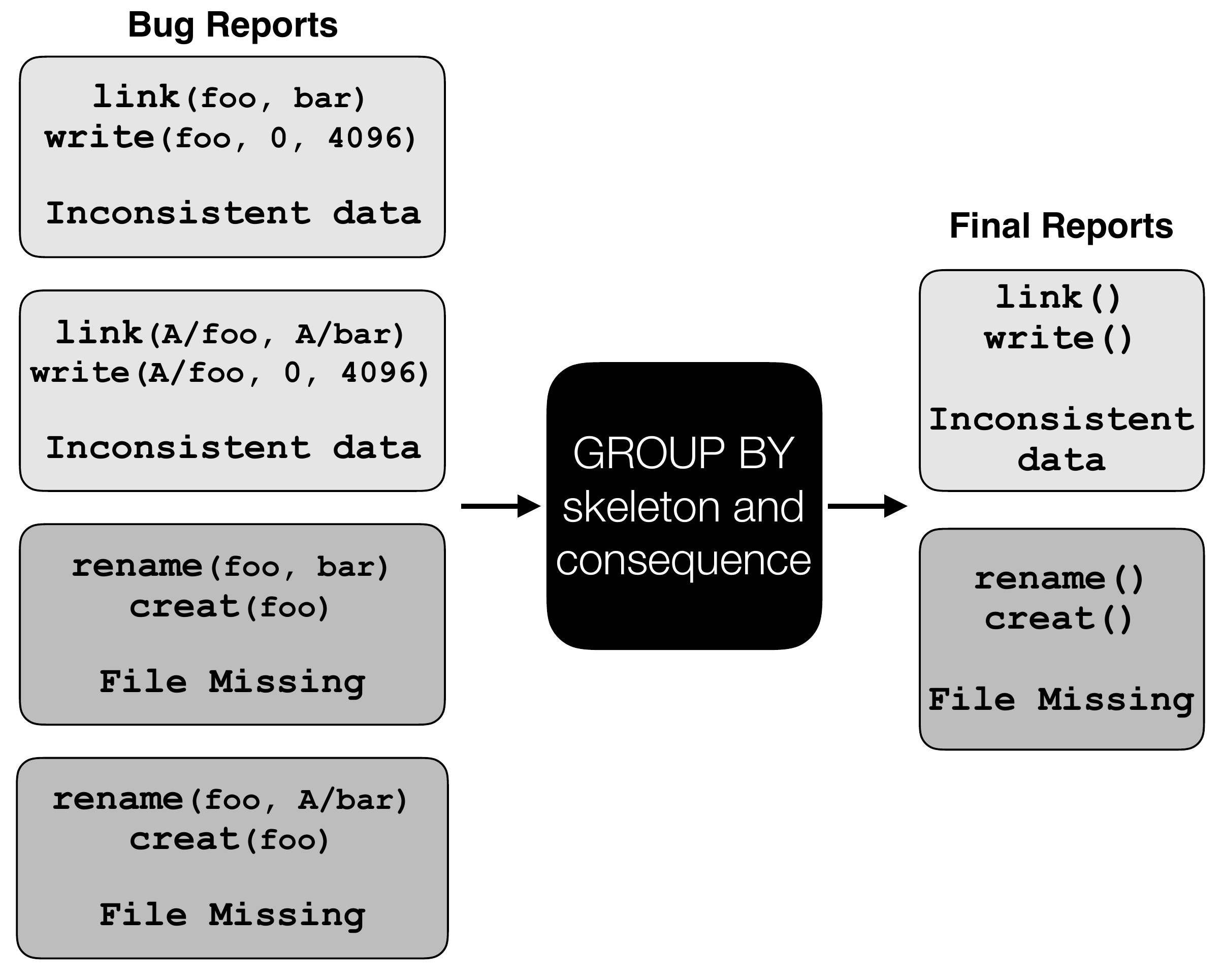}
  \mycaption{Post-processing}{The figure shows how generated bug
    reports are processed to eliminate duplicates. }
  \label{fig-postproc}
\end{figure}

%% file: eval2.tex
\section{Evaluation}
\label{sec-eval}

We evaluate the utility and performance of the \apname approach by
answering the following questions:
\begin{compactitem}
\item Do \twotools find known bugs and new bugs in Linux file systems
  in a reasonable period of time? 
  (\sref{sec-bugs})
\item What is the performance of \lsys? (\sref{sec-cm-perf})
\item What is the performance of \usys? (\sref{sec-ace-perf})
\item How much memory and CPU does \lsys consume?  (\sref{sec-consumption})
\end{compactitem}  

\input{tbl-workloads-per-seq2}

\subsection{Experimental Setup}
\label{sec-exp-setup}

\apname requires testing a large number of workloads in a systematic
manner. To accomplish this testing, we deploy \lsys on
Chameleon Cloud~\cite{mambretti2015next}, an experimental testbed for
large-scale computation.

We employ a cluster of 65 nodes on Chameleon Cloud. Each node has 24
cores, 128 GB RAM, and 250GB HDD. We install 12 VirtualBox virtual
machines running Ubuntu 16.04 LTS on each node, each with 2 GB RAM and
10 GB storage. Each virtual machine runs one instance of \lsys.
% and can be running a different kernel version from the other virtual machines in the cluster. 
Thus, we have a total of 780 virtual machines testing
workloads with \lsys in parallel. Based on the number
of virtual machines that could be reliably supported
per node, we limited the total to 780.

On a local server, we generate the workloads with \usys and divide
them into sets of workloads to be tested on each virtual machine. We
then copy the workloads over the network to each physical Chameleon
node, and, from each node, copy them to the virtual machines.

\subsection{Bug Finding}
\label{sec-bugs}

\vheading{Determining Workloads}. Our goal was to test whether the
\apname approach was useful and practical, not to exhaustively find
every crash-consistency bug. Therefore, we wanted to limit the
computational time spent on testing to a few days. Thus, we needed to
determine what workloads to test with our computational budget.

Our study of crash-consistency bugs indicated that it would be useful
to test small workloads of length one, two, and three.  However, we
estimated that testing all 25 million possible workloads of length
three was infeasible within our target time-frame. We had to further
restrict the set of workloads that we tested. We used our study to
guide us in this task. At a minimum, we wanted to select bounds that
would generate the workloads that reproduced the reported bugs. Using
this as a guideline, we came up with a set of workloads that was broad
enough to reproduce existing bugs (and potentially find new bugs), but
small enough that we could test the workloads in a few days on our
research cluster.

\vheading{Workloads}. We test workloads of length one (\seqone), two
(\seqtwo), and three (\seqthree). We further separate workloads of
length three into three groups: one focusing on data operations
(\dseqthree), one focusing on metadata operations (\mseqthree), and
one focusing on metadata operations involving a file at depth three
(\nseqthree) (by default, we use depth two).

The \seqone and \seqtwo workloads use a set of 14 file-system
operations.  For \seqthree workloads, we narrow down the list of
operations, based on what category the workload is in.  The complete
list of file-system operations tested in each category is shown in
Table~\ref{tbl-workloads-per-seq}.

\vheading{Testing Strategy}. We tested \seqone and \seqtwo workloads
on ext4, xfs, F2FS, and btrfs, but did not find any new bugs in ext4
or xfs. We additionally tested the \seqone workloads on two verified file 
systems, FSCQ and Yxv6. We focused on F2FS and btrfs for the larger \seqthree
workloads.  In total, we spend 48 hours testing all 3.37 million
workloads per file system on the 65-node research cluster described
earlier.  Table~\ref{tbl-workloads-per-seq} presents the number of
workloads in each set, and the time taken to test them (for each file
system).  All the tests are run only on 4.16 kernel. To reproduce
reported bugs, we employ the following strategy. We encode the
workload that triggers previously reported bugs in \usys. In the
course of workload generation, when \usys generates a workload
identical to the encoded one, it is added to a list. This list of
workloads is run on the kernel versions reported in
Table~\ref{tbl-bugs}, to validate that the workload produced by \usys
can indeed reproduce the bug.

\vheading{Cost of Computation}. We believe the amount of computational
effort required to find crash-consistency bugs with \twotools is
reasonable. For example, if we were to rent 780 \texttt{t2.small}
instances on Amazon to run \usys and \lsys for 48 hours, at the
current rate of \$0.023 per hour for on-demand
instances~\cite{amazon-pricing}, it would cost $780 * 48 * 0.023 =
\$861.12$. For the complete 25M workload set, the cost of computation
would go up by $7.5\times$, totaling $\$6.4$K. Thus, we can test
each file system for less than $\$7$K. Alternatively, a company can
provision physical nodes to run the tests; we believe this would not
be hard for a large company.

\vheading{Results}. \twotools found 10 \textbf{new} crash-consistency
bugs~\cite{cm-new} in btrfs and F2FS and 1 new bug in FSCQ, 
in addition to reproducing 24
out of 26 bugs reported over the past five years. We studied the bug
reports for the new bugs to ensure they were unique and not different
manifestations of the same underlying bug. We verified each unique
bug triggers a different code path in the kernel, indicating the root
cause of each bug is not the same underlying code.

All new bugs were reported to file-system developers and acknowledged
~\cite{reported1, reported2, reported3, reported4}.  Developers have
submitted patches for five bugs~\cite{patch1, patch2, patch3, patch4, patch5},
and are working on patches for the others~\cite{topatch1}.
Table~\ref{tbl-new-bugs} presents the new bugs discovered by
\twotools. We make several observations based on these results.

\input{tbl-new-bugs}

\vheading{The discovered bugs have severe consequences}. The newly
discovered bugs result in either data loss (due to missing files or
directories) or file-system corruption. More importantly, the missing
files and directories have been \emph{explicitly persisted} with an
\sysfsync call and thus should survive crashes.

\vheading{Small workloads are sufficient to reveal new bugs}. One might
expect only workloads with two or more file-system operations to
expose bugs. However, the results show that even workloads consisting
of a single file-system operation, if tested systematically, can
reveal bugs. For example, three bugs were found by \seqone workloads,
where \twotools only tested 300 workloads in a systematic fashion.
Interestingly, variants of these bugs have been patched previously,
and it was sufficient to simply change parameters to file-system
operations to trigger the same bug through a different code-path.

An F2FS bug found by \twotools is a good example of finding variants
of previously patched bugs. The previously patched bug manifested when
\sysfalloc was used with the \kpsize flag; this allocates blocks to a
file but does not increase the file size. By calling \sysfalloc with
the \kpsize flag, developers found that F2FS only checked the file
size to see if a file had been updated. Thus, \sysfdsync on the file
would have no result. After a crash, the file recovered to an
incorrect size, thereby not respecting the \kpsize flag. This bug was
patched in Nov 2017~\cite{f2fs-generic392}; however, the \sysfalloc
system call has several more flags like \fzero, \fpunch, \etc, and
developers failed to systematically test all possible parameter
options of the system call. Therefore, our tools identified and
reported that the same bug can appear when \fzero is used. Though this
bug was recently patched by developers, it provides more evidence that
the state of crash-consistency testing today is insufficient, and that
systematic testing is required.

\vheading{Crash-consistency bugs are hard to find manually}. \twotools
found eight new bugs in btrfs in kernel 4.16. Interestingly, seven of
these bugs have been present since kernel 3.13, which was released in
2014. The ability of our tools to find \emph{four-year-old} crash-consistency
bugs within two days of testing on a research cluster of modest size
speaks to both the difficulty of manually finding these bugs, and the
power of systematic approaches like \apname.

\vheading{Broken rename atomicity bug}. \usys generated several workloads that
broke the rename atomicity of btrfs. The workloads consist of first
creating and persisting a file such as \texttt{A/bar}. Next, the workload
creates another file \texttt{B/bar}, and tries to replace the original
file, \texttt{A/bar}, with the new file. The expectation is that we are
able to read either the original file, \texttt{A/bar}, or the new
file, \texttt{B/bar}. However, btrfs can lose both \texttt{A/bar} and
\texttt{B/bar} if it crashes at the wrong time. While losing rename
atomicity is bad, the most interesting part of this bug is that
\sysfsync must be called on an un-related sibling file, like \texttt{A/foo},
before the crash. This shows that workloads revealing
crash-consistency bugs are hard for a developer to find manually since
they don't always involve obvious sequences of operations.

\vheading{Crash-consistency bugs in verified file systems}. \twotools found a
crash-consistency bug in FSCQ that led to data loss in spite of 
persisting the file using \sysfdsync. The developers have acknowledged and 
patched the bug~\cite{patch5}. The origin of this bug can be tracked down 
to an optimization introduced in the C-Haskell binding in FSCQ, which is unverified 
code. 

%\vheading{Finding injected bugs}. We modified the xfs file system in
%the 4.16 kernel series to inject a bug into the rename code path. In
%this modified version of xfs, renaming a file in a directory to an
%existing file in a different directory would cause both files to
%disappear on a crash. We injected this bug by removing two lines that
%controlled the logging for the new parent inode of the renamed file.
%The in-memory version of the rename would execute correctly, but this
%change would cause data loss if a crash occurred. \usys was able to
%find our injected bug in the \seqthree workload. This synthetic bug
%shows that \twotools are capable of finding bugs for other file
%systems besides ext4, btrfs, and F2FS.

\subsection{CrashMonkey Performance}
\label{sec-cm-perf}

\lsys has three phases of operation: profiling the given workload,
constructing crash states, and testing crash-consistency.  Given a
workload, the end-to-end latency to generate a bug report is 4.6
seconds.  The main bottleneck is the kernel itself: mounting a file
system requires up-to a second of delay (if \lsys checks file-system
state earlier, it sometimes gets an error). Similarly, once the
workload is done, we also wait for two seconds to ensure the storage
subsystem has processed the writes, and that we can unmount the file
system without affecting the writes. These delays
account for 84\% of the time spent profiling.

After profiling, constructing crash states is relatively fast: \lsys
only requires 20 ms to construct each crash state. Furthermore, since
\lsys uses fine-grained correctness tests, checking crash consistency
with both read and write tests takes only 20 ms.

\subsection{Ace Performance}
\label{sec-ace-perf}

\usys generated all the workloads that were tested (3.37M) in 374
minutes ($\approx$ 150 workloads generated per second). Despite this
high cost, it is important to note that generating workloads is a
one-time cost. Once the workloads are generated, \lsys can test these
workloads on different file systems without any reconfiguration.

Deploying these workloads to the 780 virtual machines on Chameleon
took 237 minutes: 34 minutes to group the workloads by virtual
machines, 199 minutes to copy workloads to the Chameleon nodes, and 4
minutes to copy workloads to the virtual machines on each
node.

These numbers reflect the time taken for a single local server to
generate and push the workloads to Chameleon. By utilizing more
servers and employing a more sophisticated strategy for generating
workloads, we could reduce the time required to generate and push
workloads.

\subsection{Resource Consumption}
\label{sec-consumption}

The total memory consumption by \lsys averaged across 10 randomly
chosen workloads and the three sequence lengths is 20.12 MB. The low
memory consumption results from the copy-on-write nature of the
wrapper block device. Since \usys{}'s workloads typically modify small
amounts of data or metadata, the modified pages are few in number,
resulting in low memory consumption. Furthermore, \lsys uses
persistent storage only for storing the workloads (480 KB per
workload). Finally, the CPU consumption of \lsys, as reported by
\texttt{top}, was negligible (less than 1 percent).

%% file: tbl-workloads-per-seq2.tex
\begin{table*}[!t]
  \small
  \centering
  \ra{1.3}
  %\begin{tabular}{@{}lc@{}R{3cm}@{}}
   \begin{tabular}{@{}lcll@{}}
    \toprule[1.2pt]
    Sequence & File-system operations tested &\# of workloads & Run time \\
    type& & & (minutes) \\
    \midrule
    seq-1 &  \rdelim\{{3}{10pt}[] \texttt{\footnotesize{creat, mkdir, falloc, buffered write, mmap, link}} \rdelim\}{3}{20pt}[] &300 & 1\\
    seq-2 & \texttt{\footnotesize{direct-IO write, unlink, rmdir, setxattr}}  &254K & 215 \\
    & \texttt{\footnotesize{removexattr, remove, unlink, truncate}}& & \\
    seq-3-data & \texttt{\footnotesize{buffered write, mmap, direct-IO write, falloc}} &120K & 102 \\
    seq-3-metadata & \texttt{\footnotesize{buffered write, link, unlink, rename}} &1.5M & 1274\\
    seq-3-nested & \texttt{\footnotesize{link, rename}} &1.5M & 1274\\
    \midrule
    Total &  &3.37M & 2866 \\
    \bottomrule[1.2pt]
  \end{tabular}
  
  \mycaption{Workloads tested}{The table shows the number of workloads
    tested in each set, along with the time taken to test these
    workloads in parallel on 65 physical machines and the file-system operations
    tested in each category. Overall, we tested
    3.37 million workloads in two days, reproducing 24 known bugs and
    finding 10 new crash-consistency bugs.}
  \label{tbl-workloads-per-seq}
\vspace{-10pt} 
\end{table*}

%% file: tbl-new-bugs.tex
\begin{table*}[!t]
  \small
  \centering
  \ra{1.3}
  \begin{tabular}{@{}rllrr@{}}
    \toprule[1.2pt]
    Bug \# & File System & Consequence & \# of ops & Bug present since \\
    \midrule
    1 & btrfs & 
    Rename atomicity broken (file disappears) & 3 & 2014 \\
    2 & btrfs &  
    Rename atomicity broken (file in both locations) & 3 & 2018 \\
    3 & btrfs & Directory not persisted by fsync* & 3 & 2014 \\
    4 & btrfs & Rename not persisted by fsync & 3 & 2014 \\
    5 & btrfs & Hard links not persisted by fsync & 2 & 2014 \\
    6 & btrfs & Directory entry missing after fsync on directory & 2 & 2014 \\
    7 & btrfs & Fsync on file does not persist all its paths & 1 & 2014 \\
    8 & btrfs & Allocated blocks lost after fsync*  & 1 & 2014 \\
    9 & F2FS & File recovers to incorrect size* & 1 & 2015 \\
    10 & F2FS & Persisted file disappears*  & 2 & 2016 \\
    11 & FSCQ & File data loss*  & 1 & 2018 \\
    \bottomrule[1.2pt]
  \end{tabular}
  \mycaption{Newly discovered bugs}{The table shows the new bugs found
    by \twotools. The bugs have severe consequences, ranging from
    losing allocated blocks to entire files and directories
    disappearing. The bugs have been present for several years in the
    kernel, showing the need for systematic testing. Note that even
    workloads with single file-system operation have resulted in
    bugs. Developers have submitted a patch for bugs marked with *.}
  \label{tbl-new-bugs}
\vspace{-10pt}  
\end{table*}

%% file: related.tex
\section{Related Work}
\label{sec-related}
%\vspace{-5pt}

\apname offers a new point in the spectrum of techniques addressing
file-system crash consistency, alongside verified file systems and
model checking. We now place \apname in the context of prior
approaches.

\vheading{Verified File Systems}. Recent work focuses on creating new,
verified file systems from a specification~\cite{fscq, tree-verify,
  verify-push}. These file systems are proven to have strong
crash-consistency guarantees. However, the techniques employed are not
useful for testing the crash consistency of existing, widely-used
Linux file systems written in low-level languages like C. The \apname
approach targets such file systems, which are not amenable to
verification.

\vheading{Formal Crash-Consistency Models}. Ferrite~\cite{ferrite}
formalizes crash-consistency models and can be used to test if a given
ordering relationship holds in a file system; however, it is hard to
determine what relationships to test. The authors used Ferrite to test
a few simple relationships such as prefix append. On the other hand,
\usys and \lsys explore a wider range of workloads, and use oracles and
developer-provided guarantees to automatically test correctness after
a crash.

\vheading{Model Checking}. \apname is closely related to in-situ model
checking approaches such as EXPLODE~\cite{YangEtAl06-Explode} and
FiSC~\cite{fisc}.  However, unlike \apname, EXPLODE and FiSC require
modifications to the buffer cache (to see all orderings of IO
requests) and changes to the file-system code to expose choice points
for efficient checking, a complex and time-consuming task. \apname
does not require changing any file-system code and it is conceptually
simpler than in-situ model checking approaches, while still being
effective at finding crash-consistency bugs.

Though the \apname approach does not have the guarantees of
verification or the power of model checking, it has the advantage of
being easy to use (due to its black-box nature), being able to
systematically test file systems (due to its exhaustive nature), and
being able to catch crash-consistency bugs occurring on mature file
systems.

\vheading{Fuzzing}. The \apname approach bears some similarity to
fuzz-testing techniques which explore inputs that will reveal bugs in
the target system. The effectiveness of fuzzers is determined by the
careful selection of uncommon inputs that would trigger exceptional
behavior. However, \apname does not randomize input selection. Neither
does it use any sophisticated strategy to select workloads to
test. Instead, \apname exhaustively generates workloads in a bounded
space, with the bounds informed by our study or provided by the
user. While there exists fuzzers to test the correctness of system
calls~\cite{drysdale2016coverage, jones2011trinity,
  nossum2016filesystem}, there seem to be no fuzzing techniques to
expose crash-consistency bugs. The effort by Nossum and
Casasnovas~\cite{nossum2016filesystem} is closest to our work, where
they generate file-system images that are likely to expose bugs during
the normal operation of the file system (non-crash-consistency bugs).

\vheading{Record and Replay Frameworks}. 
\lsys is similar to prior
record-and-replay frameworks such as \dmlog~\cite{dmlogwrites},
Block Order Breaker~\cite{ThanuPillaiEtAl14-OSDI}, and work by Zheng
\etal~\cite{tort}. Unlike \dmlog, which requires manual
correctness tests or running \fsck, \lsys is able to automatically
test crash-consistency in an efficient manner.

Similar to \lsys, the Block Order Breaker
(BOB)~\cite{ThanuPillaiEtAl14-OSDI} also creates crash states from
recorded IO. However, BOB is only used to show that different file
systems persist file-system operations in significantly different
ways. The Application-Level Intelligent Crash Explorer (ALICE),
explores application-level crash vulnerabilities in databases, key
value stores \etc The major drawback with ALICE and BOB is that they
require the user to handcraft workloads and provide an appropriate
checker for each workload. They lack systematic exploration of the
workload space and do not understand persistence points, making it is
extremely hard for a user to write bug-triggering workloads manually.

The logging and replay framework from Zheng \etal~\cite{tort} is
focused on testing whether databases provide ACID guarantees, works
only on iSCSI disks, and uses only four workloads. \lsys is able to
test millions of workloads, and \usys allows us to generate a much
wider ranger of workloads to test.

We previewed the ideas behind \lsys in a workshop
paper~\cite{martinez2017crashmonkey}.  Since then, several features
have been added to \lsys with the prominent one being automatic
crash-consistency testing.

%% file: conc.tex
\section{Conclusion}
%\vspace{-5pt}
This paper presents Bounded Black-Box Crash Testing (\apname), a new
approach to testing file-system crash consistency. We study 26
crash-consistency bugs reported in Linux file systems over the past
five years and find that most reported bugs could be exposed by
testing small workloads in a systematic fashion. We exploit this
insight to build two tools, \twotools, that
systematically test crash consistency. Running
for two days on a research cluster of 65 machines, \twotools
reproduced 24 known bugs and found 10 new bugs in
widely-used Linux file systems.

We have made \twotools available (with demo, documentation, and a 
single line command to run \seqone workloads) at 
\url{https://github.com/utsaslab/crashmonkey}. We
encourage developers and researchers to test their file systems
against the workloads included in the repository. 

%% file: ack.tex
\section*{Acknowledgments}
We would like to thank our shepherd, Angela Demke Brown, the anonymous
reviewers, and the members of Systems and Storage Lab and LASR group
for their feedback and guidance. We would like to thank Sonika Garg,
Subrat Mainali, and Fabio Domingues for their contributions to the
CrashMonkey codebase. We are thankful to Amir Goldstein and Ted
Ts'o who encouraged us in doing this work. We also thank the Chameleon Cloud 
team for providing a research cluster to test the workloads using \lsys.
This work was supported by generous donations
from VMware, Google, and Facebook. Any opinions, findings, and
conclusions, or recommendations expressed herein are those of the
authors and do not reflect the views of other institutions.

%% file: appendix.tex
\section{Appendix}
%\vspace{-5pt}

\subsection{Bugs reproduced by \twotools}

26 of the 28 known bugs are reproducible by \twotools.
We present the workload that triggers each bug, along with
the file system on which it occurs, the difference in expected and actual
states after a crash and the overall consequence of
the bug.
\input{tbl-rep1}

\subsection{New bugs found by \twotools}
\twotools found 10 new bugs across widely used Linux file systems, btrfs and F2FS.
Additionally, our tools also found a data loss in FSCQ, ascertaining the fact that unverified components of a verified file system could lead to bugs in the final artifact.
\input{tbl-new-bugs-list}

%% file: tbl-rep1.tex
\renewcommand{\arraystretch}{1.4}
\begin{table}[H]
\begin{tabular}{|P{3cm} | P{1.8cm} | P{2.2cm} | }
\hline
Workload 1 & \multicolumn{2}{c|}{Details~\cite{btrfs-bug1}}\\
\cline{1-3}
\multirow{4}{6cm}{\Centerstack{mkdir A  \\write(0-16K) A/foo \\ sync \\ mv A/foo A/bar \\ write (0-4K) A/foo \\ fsync A/foo \\ ---Crash---  }} 
& \textbf{File system} & btrfs, F2FS\\
	\cline{2-3}
	& \textbf{Expected} & \Centerstack{  A/foo (4K) \\ A/bar (16K)}\\
	\cline{2-3}
	& \textbf{Actual} & A/foo (4K) \\
	\cline{2-3}
	& \textbf{Consequence} & Persisted file missing \\
\hline
\end{tabular}
\end{table}

\vspace{-2em}

\begin{table}[H]
\begin{tabular}{|P{3cm} | P{1.8cm} | P{2.2cm} | }
\hline
Workload 2 & \multicolumn{2}{c|}{Details~\cite{bug2}}\\
\cline{1-3}
\multirow{4}{6cm}{\Centerstack{write (0-8K) foo \\ fsync foo \\ falloc -k (8-16K) foo\\ fdatasync foo  \\ ---Crash---  }} 
& \textbf{File system} & ext4, F2FS\\
	\cline{2-3}
	& \textbf{Expected} & foo:  32 sectors\\
	\cline{2-3}
	& \textbf{Actual} & foo: 16 sectors\\
	\cline{2-3}
	& \textbf{Consequence} & Blocks allocated beyond EOF are lost\\
\hline
\end{tabular}
\end{table}

\vspace{-2em}
\renewcommand{\arraystretch}{1.7}
\begin{table}[H]
\begin{tabular}{|P{3cm} | P{1.8cm} | P{2.2cm} | }
\hline
Workload 3 & \multicolumn{2}{c|}{Details~\cite{bug3}}\\
\cline{1-3}
\multirow{4}{6cm}{\Centerstack{mkdir A \\ mkfifo A/foo \\ touch A/dummy \\ fsync A/dummy \\ mv A/foo A/bar \\ link A/bar A/foo \\ remove A/dummy \\ fsync A/dummy  \\ ---Crash---  }} 
& \textbf{File system} & btrfs\\
	\cline{2-3}
	& \textbf{Expected} & \Centerstack{A/foo \\ A/bar}\\
	\cline{2-3}
	& \textbf{Actual} & File system unmountable\\
	\cline{2-3}
	& \textbf{Consequence} & File system unmountable\\
\hline
\end{tabular}
\end{table}

\vspace{-2em}
\renewcommand{\arraystretch}{1.4}
\begin{table}[H]
\begin{tabular}{|P{3cm} | P{1.8cm} | P{2.2cm} | }
\hline
Workload 4 & \multicolumn{2}{c|}{Details~\cite{ext4-direct-write}}\\
\cline{1-3}
\multirow{4}{6cm}{\Centerstack{write (16-20K) foo \\d-write (0-4K) foo \\ ---Crash---  }} 
& \textbf{File system} & ext4\\
	\cline{2-3}
	& \textbf{Expected} & foo: Size 4K\\
	\cline{2-3}
	& \textbf{Actual} & foo: Size 0\\
	\cline{2-3}
	& \textbf{Consequence} & File metadata inconsistent\\
\hline
\end{tabular}
\end{table}

\vspace{-2em}
\renewcommand{\arraystretch}{1.4}
\begin{table}[H]
\begin{tabular}{|P{3cm} | P{1.8cm} | P{2.2cm} | }
\hline
Workload 5 & \multicolumn{2}{c|}{Details~\cite{bug5}}\\
\cline{1-3}
\multirow{4}{6cm}{\Centerstack{mkdir A \\touch A/foo\\ link A/foo A/bar\\ sync\\ unlink A/bar\\ touch A/bar\\ fsync A/bar \\ ---Crash---  }} 
& \textbf{File system} & btrfs\\
	\cline{2-3}
	&\textbf{Expected} & \Centerstack{A/foo \\ A/bar}\\
	\cline{2-3}
	& \textbf{Actual} & File system unmountable\\
	\cline{2-3}
	& \textbf{Consequence} & File system unmountable\\
\hline
\end{tabular}
\end{table}

\vspace{-2em}
\renewcommand{\arraystretch}{1.4}
\begin{table}[H]
\begin{tabular}{|P{3cm} | P{1.8cm} | P{2.2cm} | }
\hline
Workload 6& \multicolumn{2}{c|}{Details~\cite{bug6}}\\
\cline{1-3}
\multirow{4}{6cm}{\Centerstack{ mkdir A\\ touch A/foo\\ fsync A/foo \\ ---Crash--- }} 
& \textbf{File system} & btrfs\\
	\cline{2-3}
	&\textbf{Expected} & Writeable file system\\
	\cline{2-3}
	& \textbf{Actual} & Cannot create files\\
	\cline{2-3}
	& \textbf{Consequence} & Unable to create new files\\
\hline
\end{tabular}
\end{table}

\vspace{-2em}
\renewcommand{\arraystretch}{1.6}
\begin{table}[H]
\begin{tabular}{|P{3cm} | P{1.8cm} | P{2.2cm} | }
\hline
Workload 7& \multicolumn{2}{c|}{Details~\cite{bug7}}\\
\cline{1-3}
\multirow{4}{6cm}{\Centerstack{mkdir A,B,C\\ touch A/foo\\ link A/foo B/foo\_link\\ touch B/bar \\ sync\\ unlink B/foo\_link \\ mv B/bar C/bar\\ fsync A/foo \\ ---Crash---  }} 
& \textbf{File system} & btrfs\\
	\cline{2-3}
	&\textbf{Expected} & \Centerstack{A/foo \\ B \\ C/bar}\\
	\cline{2-3}
	& \textbf{Actual} & C/bar and B/bar missing\\
	\cline{2-3}
	& \textbf{Consequence} & Persisted file missing\\
\hline
\end{tabular}
\end{table}

\vspace{-2em}
\renewcommand{\arraystretch}{1.6}
\begin{table}[H]
\begin{tabular}{|P{3cm} | P{1.8cm} | P{2.2cm} | }
\hline
Workload 8& \multicolumn{2}{c|}{Details~\cite{bug8}}\\
\cline{1-3}
\multirow{4}{6cm}{\Centerstack{mkdir -p A/B \\ mkdir A/C \\ touch A/B/foo \\ touch A/B/bar \\ sync\\ mv A/B A/C\\ mkdir A/B \\ fsync A/B \\ ---Crash---  }} 
& \textbf{File system} & btrfs\\
	\cline{2-3}
	&\textbf{Expected} & \Centerstack{A/B \\ A/C/foo \\ A/C/bar}\\
	\cline{2-3}
	& \textbf{Actual} & A/B\\
	\cline{2-3}
	& \textbf{Consequence} & Renamed directory and its contents missing\\
\hline
\end{tabular}
\end{table}

\vspace{-2em}
\renewcommand{\arraystretch}{1.8}
\begin{table}[H]
\begin{tabular}{|P{3cm} | P{1.8cm} | P{2.2cm} | }
\hline
Workload 9& \multicolumn{2}{c|}{Details~\cite{btrfs-generic343}}\\
\cline{1-3}
\multirow{4}{6cm}{\Centerstack{mkdir A,B\\ touch A/foo\\ mkdir B/C\\ touch B/baz\\ sync\\ link A/foo A/bar\\ mv B/C A/ \\ mv B/baz A/\\ fsync A/foo \\ ---Crash---  }} 
& \textbf{File system} & btrfs\\
	\cline{2-3}
	&\textbf{Expected} & \Centerstack{A/C or B/C \\ A/baz or B/baz}\\
	\cline{2-3}
	& \textbf{Actual} &  \Centerstack{A/C and B/C \\ A/baz and B/baz}\\
	\cline{2-3}
	& \textbf{Consequence} & Rename persists files in both directories.\\
\hline
\end{tabular}
\end{table}

\vspace{-2em}
\renewcommand{\arraystretch}{1.4}
\begin{table}[H]
\begin{tabular}{|P{3cm} | P{1.8cm} | P{2.2cm} | }
\hline
Workload 10& \multicolumn{2}{c|}{Details~\cite{bug10}}\\
\cline{1-3}
\multirow{4}{6cm}{\Centerstack{mkdir A\\ sync\\ symlink foo, A/bar\\ fsync A \\ ---Crash---  }} 
& \textbf{File system} & btrfs\\
	\cline{2-3}
	&\textbf{Expected} & A/bar must point to foo\\
	\cline{2-3}
	& \textbf{Actual} &  A/bar is empty\\
	\cline{2-3}
	& \textbf{Consequence} & Empty symlink.\\
\hline
\end{tabular}
\end{table}

\vspace{-2em}
\renewcommand{\arraystretch}{1.8}
\begin{table}[H]
\begin{tabular}{|P{3cm} | P{1.8cm} | P{2.2cm} | }
\hline
Workload 11& \multicolumn{2}{c|}{Details~\cite{bug11}}\\
\cline{1-3}
\multirow{4}{6cm}{\Centerstack{mkdir A\\ touch A/foo\\ fsync A\\ fsync A/foo \\ mv A/foo A/bar \\ touch A/foo \\ fsync A/bar\\ ---Crash---  }} 
& \textbf{File system} & btrfs\\
	\cline{2-3}
	&\textbf{Expected} & \Centerstack{A/foo\\ A/bar}\\
	\cline{2-3}
	& \textbf{Actual} &  A/bar\\
	\cline{2-3}
	& \textbf{Consequence} & Persisted file missing\\
\hline
\end{tabular}
\end{table}

\vspace{-2em}
\renewcommand{\arraystretch}{1.4}
\begin{table}[H]
\begin{tabular}{|P{3cm} | P{1.8cm} | P{2.2cm} | }
\hline
Workload 12& \multicolumn{2}{c|}{Details~\cite{bug12}}\\
\cline{1-3}
\multirow{4}{6cm}{\Centerstack{ write (0-132K) foo\\ punch\_hole (96-128K) \\ punch\_hole (64-192K) \\ punch\_hole (32-128K) \\ fsync foo\\ ---Crash---  }} 
& \textbf{File system} & btrfs\\
	\cline{2-3}
	&\textbf{Expected} & Hole in file foo from 32-192K\\
	\cline{2-3}
	& \textbf{Actual} & Hole in file foo from 32-128K \\
	\cline{2-3}
	& \textbf{Consequence} & Extent map not persisted\\
\hline
\end{tabular}
\end{table}

\vspace{-2em}
\renewcommand{\arraystretch}{1.4}
\begin{table}[H]
\begin{tabular}{|P{3cm} | P{1.8cm} | P{2.2cm} | }
\hline
Workload 13& \multicolumn{2}{c|}{Details~\cite{btrfs-generic104}}\\
\cline{1-3}
\multirow{4}{6cm}{\Centerstack{mkdir A\\ touch A/foo\\touch A/bar\\sync\\ link A/foo A/foo\_link\\ link A/bar A/bar\_link\\ fsync A/bar\\ ---Crash---  }} 
& \textbf{File system} & btrfs\\
	\cline{2-3}
	&\textbf{Expected} & Writeable file system\\
	\cline{2-3}
	& \textbf{Actual} &  Dir A un-removable\\
	\cline{2-3}
	& \textbf{Consequence} & Directory un-removable due to incorrect entries\\
\hline
\end{tabular}
\end{table}

\vspace{-2em}
\renewcommand{\arraystretch}{1.4}
\begin{table}[H]
\begin{tabular}{|P{3cm} | P{1.8cm} | P{2.2cm} | }
\hline
Workload 14& \multicolumn{2}{c|}{Details~\cite{bug14}}\\
\cline{1-3}
\multirow{4}{6cm}{\Centerstack{touch foo\\ write (0-256K) foo\\ sync\\ mmap (0-256K) foo\\ m-write (0-4K)\\ m-write (252-256K)\\ msync (0-64K) \\ msync(192-256K)\\ ---Crash---  }} 
& \textbf{File system} & btrfs\\
	\cline{2-3}
	&\textbf{Expected} & Both mmap writes should persist\\
	\cline{2-3}
	& \textbf{Actual} &  Second mmapwrite not persisted\\
	\cline{2-3}
	& \textbf{Consequence} & Data not persisted\\
\hline
\end{tabular}
\end{table}

\vspace{-2em}
\renewcommand{\arraystretch}{1.4}
\begin{table}[H]
\begin{tabular}{|P{3cm} | P{1.8cm} | P{2.2cm} | }
\hline
Workload 15& \multicolumn{2}{c|}{Details~\cite{btrfs-generic-unremovable}}\\
\cline{1-3}
\multirow{4}{6cm}{\Centerstack{mkdir A\\sync\\ touch A/foo\\ link A/foo A/bar\\sync\\remove A/bar\\ fsync A/foo\\ ---Crash---  }} 
& \textbf{File system} & btrfs\\
	\cline{2-3}
	&\textbf{Expected} & Writeable file system\\
	\cline{2-3}
	& \textbf{Actual} &  Dir A un-removable\\
	\cline{2-3}
	& \textbf{Consequence} & Directory un-removable due to incorrect entries\\
\hline
\end{tabular}
\end{table}

\vspace{-2em}
\renewcommand{\arraystretch}{1.8}
\begin{table}[H]
\begin{tabular}{|P{3cm} | P{1.8cm} | P{2.2cm} | }
\hline
Workload 16& \multicolumn{2}{c|}{Details~\cite{btrfs-generic56}}\\
\cline{1-3}
\multirow{4}{6cm}{\Centerstack{mkdir A\\touch A/foo\\sync\\ write (0-16K) A/foo\\ fsync A/foo\\ link A/foo A/bar\\ ---Crash---  }} 
& \textbf{File system} & btrfs\\
	\cline{2-3}
	&\textbf{Expected} & foo: Size 16K\\
	\cline{2-3}
	& \textbf{Actual} &  foo: Size 0\\
	\cline{2-3}
	& \textbf{Consequence} & Data loss\\
\hline
\end{tabular}
\end{table}

\vspace{-2em}
\renewcommand{\arraystretch}{1.4}
\begin{table}[H]
\begin{tabular}{|P{3cm} | P{1.8cm} | P{2.2cm} | }
\hline
Workload 17& \multicolumn{2}{c|}{Details~\cite{bug17}}\\
\cline{1-3}
\multirow{4}{6cm}{\Centerstack{write(0-16K) foo\\fsync foo\\ sync\\ punch\_hole -k \\(8000-12096) foo\\ fsync foo\\ ---Crash---  }} 
& \textbf{File system} & btrfs\\
	\cline{2-3}
	&\textbf{Expected} & Hole must persist\\
	\cline{2-3}
	& \textbf{Actual} &  Hole not persisted\\
	\cline{2-3}
	& \textbf{Consequence} & Punch\_hole of partial page does not persist\\
\hline
\end{tabular}
\end{table}

\vspace{-2em}
\renewcommand{\arraystretch}{1.4}
\begin{table}[H]
\begin{tabular}{|P{3cm} | P{1.8cm} | P{2.2cm} | }
\hline
Workload 18& \multicolumn{2}{c|}{Details~\cite{bug18}}\\
\cline{1-3}
\multirow{4}{6cm}{\Centerstack{touch foo\\ setxattr foo u1 val1\\ setxattr foo u2, val2\\ setxattr foo u3 val3\\ sync\\ removexattr foo u2\\fsync foo\\ ---Crash---  }} 
& \textbf{File system} & btrfs\\
	\cline{2-3}
	&\textbf{Expected} & \Centerstack{u1\\u2}\\
	\cline{2-3}
	& \textbf{Actual} &  \Centerstack{u1\\u2\\u3}\\
	\cline{2-3}
	& \textbf{Consequence} & Removexattr does not persist\\
\hline
\end{tabular}
\end{table}

\vspace{-2em}
\renewcommand{\arraystretch}{1.4}
\begin{table}[H]
\begin{tabular}{|P{3cm} | P{1.8cm} | P{2.2cm} | }
\hline
Workload 19& \multicolumn{2}{c|}{Details~\cite{bug19}}\\
\cline{1-3}
\multirow{4}{6cm}{\Centerstack{mkdir A\\ touch A/foo\\sync\\ link A/foo A/bar1\\link A/foo A/bar2\\sync\\unlink A/bar2\\fsync A/foo\\ ---Crash---  }} 
& \textbf{File system} & btrfs\\
	\cline{2-3}
	&\textbf{Expected} & Writeable file system\\
	\cline{2-3}
	& \textbf{Actual} &  Dir A un-removable\\
	\cline{2-3}
	& \textbf{Consequence} & Directory un-removable due to incorrect entries\\
\hline
\end{tabular}
\end{table}

\vspace{-2em}
\renewcommand{\arraystretch}{1.4}
\begin{table}[H]
\begin{tabular}{|P{3cm} | P{1.8cm} | P{2.2cm} | }
\hline
Workload 20& \multicolumn{2}{c|}{Details~\cite{bug20}}\\
\cline{1-3}
\multirow{4}{6cm}{\Centerstack{mkdir -p A/B, C\\ touch A/B/foo\\ sync\\ mv A/B/foo C/foo\\ touch A/bar\\fsync A \\ ---Crash---  }} 
& \textbf{File system} & btrfs\\
	\cline{2-3}
	&\textbf{Expected} & \Centerstack{A/bar\\C/foo}\\
	\cline{2-3}
	& \textbf{Actual} &  \Centerstack{A/bar\\A/B/foo}\\
	\cline{2-3}
	& \textbf{Consequence} & Renamed file missing\\
\hline
\end{tabular}
\end{table}

\vspace{-2em}
\renewcommand{\arraystretch}{1.4}
\begin{table}[H]
\begin{tabular}{|P{3cm} | P{1.8cm} | P{2.2cm} | }
\hline
Workload 21& \multicolumn{2}{c|}{Details~\cite{btrfs-generic34}}\\
\cline{1-3}
\multirow{4}{6cm}{\Centerstack{mkdir A\\ touch A/foo\\ sync\\ touch A/bar\\ fsync A\\ fsync A/bar \\ ---Crash---  }} 
& \textbf{File system} & btrfs\\
	\cline{2-3}
	&\textbf{Expected} & Writeable file system\\
	\cline{2-3}
	& \textbf{Actual} & Dir A un-removable\\
	\cline{2-3}
	& \textbf{Consequence} & Directory un-removable due to incorrect entries\\
\hline
\end{tabular}
\end{table}

\vspace{-2em}
\renewcommand{\arraystretch}{1.4}
\begin{table}[H]
\begin{tabular}{|P{3cm} | P{1.8cm} | P{2.2cm} | }
\hline
Workload 22& \multicolumn{2}{c|}{Details~\cite{bug22}}\\
\cline{1-3}
\multirow{4}{6cm}{\Centerstack{touch A/foo\\ write (0-4K) A/foo\\ sync\\ mv A/foo A/bar\\fsync A/bar\\ ---Crash---  }} 
& \textbf{File system} & btrfs\\
	\cline{2-3}
	&\textbf{Expected} & A/bar\\
	\cline{2-3}
	& \textbf{Actual} &  A/foo\\
	\cline{2-3}
	& \textbf{Consequence} & Persisted file missing\\
\hline
\end{tabular}
\end{table}

\vspace{-2em}
\renewcommand{\arraystretch}{1.8}
\begin{table}[H]
\begin{tabular}{|P{3cm} | P{1.8cm} | P{2.2cm} | }
\hline
Workload 23& \multicolumn{2}{c|}{Details~\cite{bug23}}\\
\cline{1-3}
\multirow{4}{6cm}{\Centerstack{write(0-32K) foo\\sync\\link foo, bar\\ sync\\ write(32-64K) foo\\ fsync foo \\ ---Crash---  }} 
& \textbf{File system} & btrfs\\
	\cline{2-3}
	&\textbf{Expected} & foo: Size 64K\\
	\cline{2-3}
	& \textbf{Actual} &  foo: Size 32K\\
	\cline{2-3}
	& \textbf{Consequence} & Data loss\\
\hline
\end{tabular}
\end{table}

\vspace{-2em}
\renewcommand{\arraystretch}{1.4}
\begin{table}[H]
\begin{tabular}{|P{3cm} | P{1.8cm} | P{2.2cm} | }
\hline
Workload 24& \multicolumn{2}{c|}{Details~\cite{bug24}}\\
\cline{1-3}
\multirow{4}{6cm}{\Centerstack{touch foo\\mkdir A\\ fsync foo\\ sync\\ mv foo A/bar\\fsync A\\fsync A/bar \\ ---Crash---  }} 
& \textbf{File system} & btrfs\\
	\cline{2-3}
	&\textbf{Expected} & Writeable file system\\
	\cline{2-3}
	& \textbf{Actual} & Dir A un-removable\\
	\cline{2-3}
	& \textbf{Consequence} & Directory un-removable due to incorrect entries\\
\hline
\end{tabular}
\end{table}

%% file: tbl-new-bugs-list.tex
\renewcommand{\arraystretch}{2.3}
\begin{table}[H]
\begin{tabular}{|P{3cm} | P{1.8cm} | P{2.2cm} | }
\hline
Workload 1 & \multicolumn{2}{c|}{Details \cite{newbug1}}\\
\cline{1-3}
\multirow{4}{6cm}{\Centerstack{mkdir A \\ touch A/bar \\ fsync A/bar \\ mkdir B \\ touch B/bar \\ rename B/bar A/bar \\ touch A/foo \\ fsync A/foo \\ fsync A \\ ---Crash--- }}
& \textbf{File system} & btrfs\\
	\cline{2-3}
	& \textbf{Expected} & \Centerstack{  A/foo \\ A/bar or B/bar }\\
	\cline{2-3}
	& \textbf{Actual} & A/foo \\
	\cline{2-3}
	& \textbf{Consequence} & Persisted file missing \\
\hline
\end{tabular}
\end{table}

\vspace{-2em}

\begin{table}[H]
\begin{tabular}{|P{3cm} | P{1.8cm} | P{2.2cm} | }
\hline
Workload 2 & \multicolumn{2}{c|}{Details \cite{topatch1}}\\
\cline{1-3}
\multirow{4}{6cm}{\Centerstack{mkdir A \\ mkdir A/C \\ rename A/C B \\ touch B/bar \\ fsync B/bar \\ rename B/bar A/bar \\ rename A B \\ fsync B/bar \\ ---Crash---  }} 
& \textbf{File system} & btrfs \\
	\cline{2-3}
	& \textbf{Expected} & B/bar \\
	\cline{2-3}
	& \textbf{Actual} & \Centerstack{  A/bar \\ B/bar } \\
	\cline{2-3}
	& \textbf{Consequence} & Rename persists the file in both directories \\
\hline
\end{tabular}
\end{table}

\vspace{-2em}
\renewcommand{\arraystretch}{1.7}
\begin{table}[H]
\begin{tabular}{|P{3cm} | P{1.8cm} | P{2.2cm} | }
\hline
Workload 3 & \multicolumn{2}{c|}{Details \cite{patch1}}\\
\cline{1-3}
\multirow{4}{6cm}{\Centerstack{mkdir A \\  mkdir B \\ mkdir A/C \\  touch B/foo \\ fsync B/foo \\ link B/foo A/C/foo \\ fsync A \\ ---Crash---  }} 
& \textbf{File system} & btrfs\\
	\cline{2-3}
	& \textbf{Expected} & \Centerstack{B/foo \\ A/C/foo}\\
	\cline{2-3}
	& \textbf{Actual} & B/foo \\
	\cline{2-3}
	& \textbf{Consequence} & Persisted directory missing \\
\hline
\end{tabular}
\end{table}

\vspace{-2em}
\renewcommand{\arraystretch}{1.6}
\begin{table}[H]
\begin{tabular}{|P{3cm} | P{1.8cm} | P{2.2cm} | }
\hline
Workload 4 & \multicolumn{2}{c|}{Details \cite{newbug4}}\\
\cline{1-3}
\multirow{4}{6cm}{\Centerstack{mkdir A \\ sync \\ rename A B \\ touch B/foo \\ fsync B/foo \\ fsync B \\ ---Crash---  }} 
& \textbf{File system} & btrfs \\
	\cline{2-3}
	& \textbf{Expected} & \Centerstack{ B \\ B/foo } \\
	\cline{2-3}
	& \textbf{Actual} & A/foo \\
	\cline{2-3}
	& \textbf{Consequence} & Persisted file missing \\
\hline
\end{tabular}
\end{table}

\vspace{-2em}
\renewcommand{\arraystretch}{1.4}
\begin{table}[H]
\begin{tabular}{|P{3cm} | P{1.8cm} | P{2.2cm} | }
\hline
Workload 5 & \multicolumn{2}{c|}{Details \cite{newbug4}}\\
\cline{1-3}
\multirow{4}{6cm}{\Centerstack{ mkdir A \\ mkdir B \\ touch A/foo \\ link (A/foo, B/foo) \\  fsync A/foo \\ fsync B/foo \\ ---Crash---  }} 
& \textbf{File system} & btrfs\\
	\cline{2-3}
	&\textbf{Expected} & \Centerstack{A/foo \\ B/foo}\\
	\cline{2-3}
	& \textbf{Actual} & A/foo \\
	\cline{2-3}
	& \textbf{Consequence} & Hard link missing even after fsync \\
\hline
\end{tabular}
\end{table}

\vspace{-2em}
\renewcommand{\arraystretch}{1.4}
\begin{table}[H]
\begin{tabular}{|P{3cm} | P{1.8cm} | P{2.2cm} | }
\hline
Workload 6& \multicolumn{2}{c|}{Details \cite{reported1}}\\
\cline{1-3}
\multirow{4}{6cm}{\Centerstack{ mkdir test \\ mkdir test/A \\ touch test/foo \\ touch test/A/foo \\ fsync test/A/foo \\ fsync test \\ ---Crash--- }} 
& \textbf{File system} & btrfs \\
	\cline{2-3}
	&\textbf{Expected} & \Centerstack{test/A/foo \\ test/foo} \\
	\cline{2-3}
	& \textbf{Actual} & test/A/foo \\
	\cline{2-3}
	& \textbf{Consequence} & File missing in spite of persisting parent directory \\
\hline
\end{tabular}
\end{table}

\vspace{-2em}
\renewcommand{\arraystretch}{1.6}
\begin{table}[H]
\begin{tabular}{|P{3cm} | P{1.8cm} | P{2.2cm} | }
\hline
Workload 7& \multicolumn{2}{c|}{Details \cite{reported4}}\\
\cline{1-3}
\multirow{4}{6cm}{\Centerstack{touch foo \\ mkdir A \\  link (foo, A/bar) \\ fsync foo \\ ---Crash---  }} 
& \textbf{File system} & btrfs\\
	\cline{2-3}
	&\textbf{Expected} & \Centerstack{foo \\ A/bar}\\
	\cline{2-3}
	& \textbf{Actual} & foo \\
	\cline{2-3}
	& \textbf{Consequence} & Fsync of a file does not persist all its names\\
\hline
\end{tabular}
\end{table}

\vspace{-2em}
\renewcommand{\arraystretch}{1.6}
\begin{table}[H]
\begin{tabular}{|P{3cm} | P{1.8cm} | P{2.2cm} | }
\hline
Workload 8& \multicolumn{2}{c|}{Details \cite{patch2}}\\
\cline{1-3}
\multirow{4}{6cm}{\Centerstack{write (0-16K) foo \\ Fsync foo \\ falloc -k (16-20K) \\ fsync(foo) \\ ---Crash---  }} 
& \textbf{File system} & btrfs\\
	\cline{2-3}
	&\textbf{Expected} & foo: 40 sectors \\
	\cline{2-3}
	& \textbf{Actual} & foo: 32 sectors \\
	\cline{2-3}
	& \textbf{Consequence} & Blocks allocated beyond EOF are lost\\
\hline
\end{tabular}
\end{table}

\vspace{-2em}
\renewcommand{\arraystretch}{1.8}
\begin{table}[H]
\begin{tabular}{|P{3cm} | P{1.8cm} | P{2.2cm} | }
\hline
Workload 9& \multicolumn{2}{c|}{Details \cite{patch3}}\\
\cline{1-3}
\multirow{4}{6cm}{\Centerstack{write (0-16K) foo \\ fsync foo \\ fzero -k (16-20K) \\  fsync(foo) \\ ---Crash---  }} 
& \textbf{File system} & F2FS \\
	\cline{2-3}
	&\textbf{Expected} & foo: Size16K \\
	\cline{2-3}
	& \textbf{Actual} & foo: Size 20K \\
	\cline{2-3}
	& \textbf{Consequence} & Recovers to incorrect file size \\
\hline
\end{tabular}
\end{table}

\vspace{-2em}
\renewcommand{\arraystretch}{1.4}
\begin{table}[H]
\begin{tabular}{|P{3cm} | P{1.8cm} | P{2.2cm} | }
\hline
Workload 10& \multicolumn{2}{c|}{Details \cite{patch4}}\\
\cline{1-3}
\multirow{4}{6cm}{\Centerstack{mkdir A \\ sync \\ rename A B \\ touch B/foo \\ fsync B/foo \\ ---Crash---  }} 
& \textbf{File system} & F2FS \\
	\cline{2-3}
	&\textbf{Expected} & B/foo \\
	\cline{2-3}
	& \textbf{Actual} &  A/foo \\
	\cline{2-3}
	& \textbf{Consequence} & Persisted file ends up in a different directory \\
\hline
\end{tabular}
\end{table}

\vspace{-2em}
\renewcommand{\arraystretch}{1.4}
\begin{table}[H]
\begin{tabular}{|P{3cm} | P{1.8cm} | P{2.2cm} | }
\hline
Workload 11& \multicolumn{2}{c|}{Details}\\
\cline{1-3}
\multirow{4}{6cm}{\Centerstack{write (0-4K) foo\\ sync\\ write (4-8K) foo\\ fdatasync foo \\ ---Crash---  }} 
& \textbf{File system} & FSCQ \\
	\cline{2-3}
	&\textbf{Expected} & foo: Size 8K \\
	\cline{2-3}
	& \textbf{Actual} &  foo: Size 4K \\
	\cline{2-3}
	& \textbf{Consequence} & Data loss \\
\hline
\end{tabular}
\end{table}